\documentclass[a4paper,12pt]{article}

\usepackage{epsfig,graphicx,amsmath,amssymb}
\usepackage{url}
\usepackage{hyperref}
\usepackage{subcaption}
\usepackage{siunitx,mathtools}
\usepackage{mhchem,booktabs}
\newcommand{\ee}{e^{+} e^{-}}

\def \ee {e^+e^-}

\newcommand{\fplus}{f^{\eta\pi}_+}
\newcommand{\fzero}{f^{\eta\pi}_0}

\newcommand{\mpid}{m_\pi^2}

\newcommand{\lambetapi}[1]{\sqrt{\lambda_{\eta\pi}{(#1)}}}

\newcommand{\fig}[1]{\ref{fig:#1}}
\newcommand{\lbl}[1]{\label{eq:#1}}
\newcommand{\rf}[1]{(\ref{eq:#1})}
\newcommand{\be}{\begin{equation}}
\newcommand{\en}{\end{equation}}
\newcommand{\dotfplus}{\dot{f}^{\eta\pi}_+}
\newcommand{\disc}{\hbox{disc}}
\newcommand{\bc}{\begin{center}}
\newcommand{\ec}{\end{center}}

\DeclareSIUnit[number-unit-product = \,]{\promille}{\textperthousand}
\DeclareSIUnit\ch{ch}
\DeclareMathOperator{\Tr}{Tr}

\newlength{\mylinewidth}
\newcommand{\Lag}{\mathcal{L}}

\newcommand{\pd}{\partial}
\newcommand{\relphantom}[1]{\mathrel{\phantom{#1}}}

\newcommand{\W}[4]{%
  \ifthenelse
  {%
    \equal{#2}{}%
  }%
  {%
    W^{#1\,\hphantom{#3}#4}_{\:\hphantom{#1}#3}%
  }%
  {%
    W^{#1\,#2}_{\:\hphantom{#1\,#2}#3}%
  }%
}

\def\Journal#1#2#3#4{{#1}~{\bf #2} (#4) #3}

\def\NPB{{\em Nucl.~Phys.}~B}
\def\PLB{{\em Phys.~Lett.}~B}

\def\PRD{{\em Phys.~Rev.}~D}

\def\EPJ{{\em Eur.~Phys.~J.}~C}

\begin{document}

\thispagestyle{empty}

$\phantom{.}$

\begin{flushright}
{\sf  MITP/14-015 \\
  } 
\end{flushright}

\hfill

\begin{center}
{\Large {\bf Light Meson Dynamics Workshop\\
Mini-proceedings} \\
\vspace{0.75cm}}

\vspace{1cm}

{\large February 10--12, 2014 in Mainz, Germany}

\vspace{2cm}

{\it Editors:}
Wolfgang Gradl, Pere Masjuan, Michael Ostrick, and Stefan Scherer
\vspace{0.5cm}

PRISMA Cluster of Excellence, Institut f\"ur Kernphysik, Johannes Gutenberg-Universit\"at Mainz, Germany\\

\vspace{2.5cm}

ABSTRACT

\end{center}

\vspace{0.3cm}

\noindent
The mini-proceedings of the Light Meson Dynamics Workshop held in Mainz from February 10$^{\rm th}$ to 12$^{\rm th}$, 2014, are presented.

\medskip\noindent
The web page of the conference, which contains all talks, can be found at
\begin{center}
\url{https://indico.cern.ch/event/287442/overview} 
\end{center}

\vspace{0.5cm}

\newpage

{$\phantom{=}$}

\vspace{0.5cm}

\tableofcontents

\newpage

\section{Introduction to the Workshop}

\addtocontents{toc}{\hspace{1cm}{\sl W.~Gradl, M.~Ostrick, and S.~Scherer}\par}

\vspace{5mm}

\noindent
 W.~Gradl, M.~Ostrick, and S.~Scherer

\vspace{5mm}

\noindent
 PRISMA Cluster of Excellence, Institut f\"ur Kernphysik, Johannes Gutenberg-Universit\"at Mainz,  D-55099 Mainz, Germany\\

\vspace{5mm}

   The strong-interaction part of the Standard Model is described by
an SU(3) gauge theory---Quantum Chromodynamics (QCD)---in terms of quarks
and gluons as the fundamental dynamical degrees of freedom.
   However, experimentally only color-neutral combinations, namely,
mesons and baryons, are observed as the asymptotic states of the theory.
   Unraveling the structure and dynamics of (light) mesons is still one of
the fascinating challenges of the strong interactions.
   The scope of the workshop was to identify and discuss key issues of the
field in combination with relevant experimental and theoretical tools.
   The subject-matter may be summarized as follows.
\begin{itemize}
\item Properties of light (and not so light) mesons

   The principal properties are masses and widths, where it is particularly
important to extract the pole parameters from experiment in a model-independent
way.
   Further structure information deals with (transition) form factors, decay rates
and distributions, etc. Last but not least, the question of the nature of a resonance
is of especial relevance.

\item Dynamics of mesons

   The dynamics of mesons reveals itself in terms of scattering, (production) cross sections,
the generation of resonances, and the response to external probes.
   The interactions of pseudoscalars with pseudoscalars (PP), pseudoscalars with pseudovectors
(PV), vectors with vectors (VV), plus additional electromagnetic interactions (PV$\gamma$, \ldots)
are key processes to be understood, in particular, in view of the generation of resonances.
   Finally, also weak interactions yield valuable information on the dynamics of light
mesons, mainly for kaons.

\item Connection to QCD

   A direct connection to QCD is provided by lattice field theory.
   In terms of effective theories, an important link is given by chiral
symmetry in terms of Ward identities.
   The interplay of three distinct symmetry-breaking mechanisms (dynamical spontaneous symmetry
breaking, explicit symmetry breaking due to the quark masses, and the U(1)
axial anomaly) generates an extremely rich physics case.
   In this context, the $\eta-\eta'$ system provides a unique stage for
studying all three mechanisms simultaneously.

\item Precision calculations

   It is compulsory to perform precision calculations for at least a few key (strong-interaction)
quantities such as, e.g., pion scattering lengths, the $\pi^0\to\gamma\gamma$ decay rate, etc.
   The inclusion of isospin-symmetry breaking in terms of different $u$- and $d$-quark masses
as well as electromagnetism plays an important role.
   Moreover, precision calculations are needed for observables relevant for essential Standard
Model predictions.


\item Rare or forbidden decays

In parallel to the hadronic decays discussed above, there are many
$C$ and $CP$ violating decay modes of $\eta$ and $\eta^\prime$ mesons.
In the next years, the huge collected data samples will allow us to
improve the upper limits on several of these branching fractions by at least one
order of magnitude.

\item Theoretical tools

   The main theoretical tools discussed at the workshop include chiral perturbation
theory, unitarized chiral dynamics, phenomenological approaches including quantum
corrections and various applications of dispersion relations.

\item  Experimental tools

The experimental study of light meson dynamics is based on exclusive 
measurements of meson decays with large-acceptance detectors. 
At the workshop recent results and perspectives of the BESIII, Crystal Ball, 
KLOE, NA48/62 and WASA experiments were presented and discussed. 
One focus is the large number of charged and neutral $\eta$, $\eta'$, $K$ and $\omega$ 
decays which are intimately related to the low-energy dynamics of QCD. The experiments 
differ in the meson production mechanisms and background conditions as
well as their capabilities in calorimetry and tracking, providing
complementary approaches to the measurements. \\
Different experiments also have access to different regions of the
electromagnetic transition form factors, allowing a quantitative
connection between the time-like and the space-like regions. \\
We expect to improve the available statistics on $K^\pm$, $\eta$,
$\eta^\prime$ and $\omega$ mesons by several orders of magnitude
within the next few years.

\end{itemize}

\vspace{2cm}

\noindent
We acknowledge the support of the Deutsche Forschungsgemeinschaft DFG through the Collaborative Research Center ``The Low-Energy Frontier of the Standard Model" (SFB 1044).

\vspace{5mm}
\noindent
This work is a part of the activity of the SFB 1044:
\begin{center}
[\url{http://sfb1044.kph.uni-mainz.de/sfb1044/}]
\end{center}

\newpage

\section{Summaries of the talks}

\subsection{The Odd Intrinsic Parity Sector of Chiral Perturbation Theory }
\addtocontents{toc}{\hspace{2cm}{\sl J.~Bijnens}\par}

\vspace{5mm}

J.~Bijnens

\vspace{5mm}

\noindent
Department of Astronomy and Theoretical Physics, Lund University,\\
S\"olvegatan 14A, SE 223 62 Lund, Sweden\\

\vspace{5mm}

The chiral anomaly started with the decay
$\pi^0\to\gamma\gamma$ \cite{Steinberger:1949wx}
and its conflict with the naive Ward identities
as solved by the discovery of the chiral anomaly
\cite{Adler:1969gk1}. At the same time current algebra and
effective Lagrangians started but no naive Lagrangian with the correct chiral
invariance was found. Ref.~\cite{Wess:1971yu} solved the problem by directly
integrating the anomalous divergence of 
\cite{Adler:1969gk1} resulting in an effective action formulated in five
dimensions. Witten\cite{Witten:1983tw} clarified the structure
of this effective
action, known as the Wess-Zumino-Witten (WZW) term.
Another aspect is intrinsic parity, which is like parity
but without its space-time part. A Lorentz-invariant Lagrangian
that has parity, is also intrinsic parity invariant except when
$\epsilon_{\mu\nu\alpha\beta}$ is present. Processes with an odd number of
pseudo-scalars thus require an $\epsilon_{\mu\nu\alpha\beta}$, hence
the close connection of odd-intrinsic-parity and the anomaly,
reviewed in e.g. \cite{Bijnens:1993xi}.

Discussion of chiral logarithms started in the 1970s.
For $\pi^0,\eta\to\gamma\gamma$
they were found to vanish \cite{Donoghue:1986wv} in agreement with the
naive expectation from the anomaly nonrenormalization
theorem.
Corrections were however found for the same process
with one off-shell photon
\cite{Bijnens:1988kx,Donoghue:1988ct}. The reason this is allowed is that
the WZW term describes the anomaly but higher order terms that are chiral
invariant can contribute to the same processes. But also loop diagrams
including the WZW vertices must be fully invariant.
This is true for the full divergence structure \cite{Bijnens:1989jb},
see also \cite{Issler:1990nj}.
The full version of the NLO Lagrangian was obtained in
\cite{Bijnens:2001bb}. The two-flavour
one including virtual photons is also known \cite{Ananthanarayan:2002kj}.

The decay $\pi^0\to\gamma\gamma$ is the  main test of the anomaly.
There are two precise experiments, PRIMEX \cite{Larin:2010kq} and CERN \cite{Atherton:1985av}. The theory has an enhancement over the anomaly predictions
resulting mainly from $\pi^0$-$\eta$-$\eta^\prime$ mixing
\cite{Ananthanarayan:2002kj,Goity:2002nn}.  Electromagnetic \cite{Ananthanarayan:2002kj} and higher loop effects are quite small
\cite{Kampf:2009tk,Bijnens:2012hf}. A recent review is \cite{Bernstein:2011bx}.
The agreement of the prediction of $8.1\pm0.1$~eV is in good agreement with the
measurement $7.82\pm0.22$~eV, a test to about 3\%

The other main test, $\pi\gamma\to\pi\pi$, is not quite as precise.
The main measurement \cite{Antipov:1986tp}
agrees satisfactorily with theory after including the
one-loop corrections \cite{Bijnens:1989ff}
and the surprisingly large electromagnetic
corrections \cite{Ametller:2001yk}.
Higher order leading logarithms \cite{Bijnens:2012hf} arte small.
The evaluation from $\pi e\to\pi\pi e$ \cite{Giller:2005uy}
has a similar agreement with the predictions, to about 10\%.

There are also anomalous form-factors in several weak decays. The precision
varies but allows for clear tests of the \emph{sign} of the
anomaly. The ChPT calculations are in \cite{Ametller:1993hg}.
There are more processes that have the anomaly. Many are treated in
other talks at this conference. e.g. $\eta,\pi^0\to\gamma^*\gamma^*$
or $\eta\to\pi^+\pi^-\gamma$ including corrections from dispersion theory.
Some oddities that allow in some domains of phase space for interesting
effects are the processes $\gamma\gamma\to3\pi$ and $\eta\to\pi\pi\gamma\gamma$,
known at
tree level \cite{Bos:1994yw} and at one-loop \cite{Talavera:1995fx}.

The NLO Lagrangian is known but only partial fits of the parameters
to experiment exist \cite{Strandberg:2003zf}.
There exists many estimates,
starting with the HLS model in \cite{Bijnens:1989jb} and the chiral quark model
\cite{Bijnens:1991db} and the full resonance saturation 
study \cite{Kampf:2011ty}.

Finally, some higher loops are known, two-loops for $\pi^0\to\gamma\gamma$
\cite{Kampf:2009tk}, partial results for
$\eta\to\gamma\gamma$ \cite{Bijnens:2010pa} and various leading logarithms up
to six loops \cite{Bijnens:2012hf}.

\newpage

\subsection{$\eta$-$\eta^\prime$ mixing: overview }
\addtocontents{toc}{\hspace{2cm}{\sl R.~Escribano}\par}

\vspace{5mm}

R.~Escribano

\vspace{5mm}

\noindent
Grup de F\'{\i}sica Te\`orica (Departament de F\'{\i}sica) and
Institut de F\'{\i}sica d'Altes Energies (IFAE),
Universitat Aut\`onoma de Barcelona,
E-08193 Bellaterra (Barcelona), Spain

\vspace{5mm}

The subject of $\eta$-$\eta^\prime$ mixing is now becoming interesting in view of the present and forthcoming experiments at
COSY (J\"ulich), DAPHNE (Frascati), ELSA (Bonn), MAMI (Mainz), VEPP-2000 (BINP, Novosibirsk), CEBAF (JLAB), BEPCII/BESIII (Beijing) and
B-factories (BABAR, Belle and Belle II)
where many different processes involving $\eta$ and/or $\eta^\prime$ mesons are/will be measured abundantly and precisely
as compared to earlier experiments.

Relevant topics concerning $\eta$-$\eta^\prime$ mixing are the mixing parameters, that is, the pseudoscalar decay constants associated with $\eta$ and $\eta^\prime$
and the related mixing angles in the octet-singlet and quark-flavour bases, the possibility of a gluonic content in the $\eta^\prime$ wave function, and
the different sets of observables
($V\to P\gamma$ decays, with $V=\rho, \omega, \phi$ and $P=\eta, \eta^\prime$, $J/\psi\to VP$ decays, and $\eta$ and $\eta^\prime$ transition form factors,
among the most precise sets)
where these parameters can be extracted from.

Concerning the mixing parameters, a brief introductory summary is the following.
There are two kinds of mixing, that of mass eigenstates and that of decay constants.
The mixing of mass eigenstates consists of a rotation matrix described in terms of single mixing angle,
$\theta_P$ in the octet-singlet basis and $\phi_P$ in the quark-flavour basis, that connects the mathematical states, $\eta_8$ and $\eta_0$ or $\eta_q$ and $\eta_s$,
depending on the basis, to the physical states $\eta$ and $\eta^\prime$.
Both mixing angles are related through $\theta_P=\phi_P-\arctan\sqrt{2}$.
In this mixing scheme three assumptions are implicit:
i) there is no mixing with other pseudoscalars ($\pi^0$, $\eta_c$, radial excitations, glueballs\ldots);
ii) the mixing angle is real (supported by the fact that $\Gamma_{\eta, \eta^\prime}\ll m_{\eta, \eta^\prime}$); and
iii) there is no energy dependence.
The mixing of decay constants is characterized by $\langle 0|A_\mu^{a(i)}|\eta^{(\prime)}(p)\rangle=i\sqrt{2}F_{\eta^{(\prime)}}^{a(i)} p_\mu$,
with $a=8,0(i=q,s)$ and $A_\mu^{a(i)}$ the corresponding axial-vector current.
The four independent decay constants can be parameterised in terms of either $F_{8,0}$, the octet and singlet decay constants, and two mixing angles $\theta_{8,0}$,
in the octet-singlet basis, or $F_{q,s}$, the light-quark and strange decay constants, and the mixing angles $\phi_{q,s}$, in the quark-flavour basis, respectively.
Are all these mixing angles related?
To answer this question, one must resort to Large-$N_c$ Chiral Perturbation Theory \cite{Kaiser:2000gs},
where the effects of the pseudoscalar singlet $\eta_0$ are treated perturbatively in a simultaneous expansion in $p^2$, $m_q$ and $1/N_c$.
In this framework, one can see:
i) that a one mixing angle scheme can only be used at leading order in this expansion, where $\theta_8=\theta_0=\theta_P$ (or $\phi_q=\phi_s=\phi_P$)
and the decay constants are equal;
ii) that at next-to-leading order the two mixing angles scheme must be used, thus making a difference between $\theta_8$ and $\theta_0$ and with respect to $\theta_P$
(or similarly between $\phi_q$ and $\phi_s$ and with respect to $\phi_P$) and where the decay constants are all different among themselves; and
iii) that the mixing structure of the decays constants and the fields is exactly the same.
For a compendium of formulae see Refs.~\cite{Feldmann:1998vh,Kaiser:1998ds,Feldmann:1998sh,Escribano:2005qq}.
At the same time, one can also see that $\sin(\theta_8-\theta_0)\propto (F_K^2-F_\pi^2)$, a $SU(3)$-breaking effect expected to be of the order of 20\% ($F_K/F_\pi\simeq 1.2$),
and $\sin(\phi_q-\phi_s)\propto\Lambda_1$, an OZI-rule breaking parameter expected to be small.
In the FKS scheme \cite{Feldmann:1999uf}, this $\Lambda_1$ parameter is assumed to be negligible,
a hypothesis that is tested experimentally since the two mixing angles are seen to be compatible \cite{Escribano:2005qq}.
If one forces this equality, $\phi_q=\phi_s=\phi_P$, which is not based in theory, the result of the fit is
$F_q/F_\pi=1.10\pm 0.03$, $F_s/F_\pi=1.66\pm 0.06$, and $\phi_P=(40.6\pm 0.9)^\circ$ \cite{Escribano:2005qq}.
Therefore, a recommendation for experimental collaborations would be to use for the time being
(until the achieved accuracy permits to distinguish between $\phi_q$ and $\phi_s$) the quark-flavour basis in their analyses.
To finish, just to mention that the decay constants $F_\eta$ and $F_{\eta^\prime}$ do not exist similarly to $F_\pi$ or $F_K$
but instead the four different decays constants mentioned before, in one basis or the other, must be used for the $\eta$-$\eta^\prime$ system.
The interested reader can use Ref.~\cite{Feldmann:1999uf} as a reference text for a complete introduction to these topics and 
a detailed list of publications and analyses prior to year 2000.

Concerning the possible gluonic content in the $\eta^\prime$ wave function,
two complete and precise sets of experimental data haven taken into account to explore this possibility:
the $V\to P\gamma$ decays, with $V=\rho, \omega, \phi$ and $P=\eta, \eta^\prime$, and the $J/\psi\to VP$ decays.
In the first case, using a very general model for $VP\gamma$ transitions \cite{Bramon:2000fr}, one gets 
$\phi_P=(41.4\pm 1.3)^\circ$ and $Z_{\eta^\prime}^2=0.04\pm 0.09$, or, equivalently, $|\phi_{\eta^\prime G}|=(12\pm 13)^\circ$
(the parameter $Z_{\eta^\prime}$ weights the amount of gluonium in the wave function and $\phi_{\eta^\prime G}=-\arcsin Z_{\eta^\prime}$),
that is, absence of gluonium in the $\eta^\prime$ \cite{Escribano:2007cd}.
This result is in contradiction with the experimental analysis performed by the KLOE Collaboration,
where, using several ratios of $V\to P\gamma$ decays, described by the same model as before, in addition to the ratio $\eta^\prime/\pi^0\to\gamma\gamma$,
they found $\phi_P=(40.4\pm 0.6)^\circ$ and $Z_{\eta^\prime}^2=0.12\pm 0.04$ \cite{Ambrosino:2009sc},
thus confirming their first analysis with the results $\phi_P=(39.7\pm 0.7)^\circ$ and $Z_{\eta^\prime}^2=0.14\pm 0.04$ \cite{Ambrosino:2006gk}.
The reason for the discrepancy between the first phenomenological analysis mentioned above and the former two experimental analyses 
is the inclusion in the latter of the ratio $\eta^\prime/\pi^0\to\gamma\gamma$ in the fits.
This sole observable makes the difference.
However, we believe that the way KLOE characterises this ratio, as a function of $F_q$, $F_s$, $\phi_P$, and, simultaneously, $Z_{\eta^\prime}$
is a contradiction in terms, since Chiral Perturbation Theory assumes that $\eta$ and $\eta^\prime$ are quark-antiquark bound states.
In the case of $J/\psi\to VP$ decays, the values obtained were $\phi_P=(44.6\pm 4.4)^\circ$ and $Z_{\eta^\prime}^2=0.29^{+0.18}_{-0.26}$ \cite{Escribano:2008rq},
thus drawing a conclusion less definitive but in accord with the $V\to P\gamma$ phenomenological analysis.
Anyway, more refined experimental data will contribute decisively to clarify this issue.
For completion, when the gluonic content of the $\eta^\prime$ is not allowed, $Z_{\eta^\prime}=0$,
the fitted value of the $\eta$-$\eta^\prime$ mixing angle in the quark-flavour basis is found to be $\phi_P=(41.5\pm 1.2)^\circ$, from $V\to P\gamma$ decays \cite{Escribano:2007cd}, 
and $\phi_P=(40.7\pm 2.3)^\circ$, from $J/\psi\to VP$ decays \cite{Escribano:2008rq}, respectively.
Other relevant analyses on this topic are Refs.~\cite{Kou:1999tt,Thomas:2007uy}.

Finally, a more recent and novel approach for the extraction of the $\eta$-$\eta^\prime$ mixing parameters
is the analysis of the $\eta$ and $\eta^\prime$ transition form factors in the space-like region at low and intermediate energies
in a model-independent way through the use of rational approximants (see P.~Masjuan's contribution to these proceedings for more details).
Using the normalization of the form factors as obtained from the experimental $\eta^{(\prime)}\to\gamma\gamma$ decay widths as well as
the fitted result for the asymptotic value of the $\eta$ form factor, 
one gets $F_q/F_\pi=1.06\pm 0.01$, $F_s/F_\pi=1.56\pm 0.24$, and $\phi_P=(40.3\pm 1.8)^\circ$ \cite{Escribano:2013kba},
in nice agreement with previous results, a bit less precise but very promising for the near future if more space- and time-like experimental data
for these form factors are released together with a more precise measurement of the decay widths.

\newpage

\subsection{$\eta$ and $\eta^\prime$ physics at BESIII}
\addtocontents{toc}{\hspace{2cm}{\sl S.~Fang}\par}

\vspace{5mm}

S.~Fang (on behalf of the BESIII collaboration)

\vspace{5mm}

\noindent
Institute of High Energy Physics, Beijing, China\\
\vspace{5mm}

Both $\eta$ and $\eta^\prime$,
discovered about half of a century ago, are two important states in the lightest pseduoscalar nonet,  which attracts considerable interest in the decays both theoretically and experimentally because of their special roles in low energy scale quantum chromodynamics 
theory.  Their dominant radiative and hadronic decays were observed and well measured, but the study of their anomalous  decays is still an open field.  A sample of 225.3 million $J/\psi$ events taken at the BESIII detector in 2009  offers a unique opportunity to study $\eta$ and $\eta^\prime$ decays via $J/\psi \rightarrow \gamma \eta(\eta^\prime)$ or $J/\psi\rightarrow\phi\eta(\eta^\prime)$.

 With a new level of precision,
the Dalitz plot parameters for $\eta^\prime\rightarrow \pi^+\pi^-\eta$ are determined in a generalized and a linear representation~\cite{bes3_dalitz}. In general the results are in reasonable agreement with the previous measurements and the C-parity violation is not evident.  The statistical error of the parameters are still quite large,
much more data strongly needed to provide a more stringer test of the chiral theory. The decays of $\eta^\prime\rightarrow \pi^+\pi^-e^+e^-$ and $\eta^\prime\rightarrow \pi^+\pi^-\mu^+\mu^-$ were also studied via $J/\psi\rightarrow \gamma\eta^\prime$~\cite{bes3_ppmm}. A clear $\eta^\prime$ peak is observed in 
the $M_{\pi^+\pi^-e^+e^-}$mass spectrum, and the branching fraction is measured to be $B(\eta^\prime\rightarrow\pi^+\pi^-e^+e^-)=(2.11 \pm 0.12\pm 0.14)
\times 10^{-3}$, which is in good agreement with theoretical predictions~\cite{theo} and the previous measurement~\cite{cleo}, but is determined with much higher precision. The mass spectra of $M_{\pi^+\pi^-}$ and $M_{e^+e^-}$ are also consistent with the theoretical predictions~\cite{theo} that $M_{\pi^+\pi^-}$ is dominated by $\rho^0$ , and $M_{e^+e^-}$ has a peak just above 2$m_e$ .   No $\eta^\prime$ signal is found in the $M_{ \pi^+\pi^-\mu^+\mu^-}$ mass spectrum, and the upper limit is determined to be 
$B(\eta^\prime\rightarrow  \pi^+\pi^-\mu^+\mu^-) < 2.9\times10^{-5} $ at the 90\% confidence level.  To test the fundamental symmetries, a search for P and CP violation decays of
$\eta/\eta^\prime\rightarrow \pi^+\pi^-,\pi^0\pi^0$ was performed~\cite{bes3_cp}.  No evident signals were observed, and then the branching fraction upper limits,
$B(\eta\rightarrow\pi^+\pi^-)<3.9\times 10^{-4}$,$B(\eta\rightarrow\pi^0\pi^0)<6.9\times 10^{-4}$,
$B(\eta^\prime\rightarrow\pi^+\pi^-)<5.5\times 10^{-5}$ and $B(\eta^\prime\rightarrow\pi^0\pi^0)<4.5\times 10^{-4}$,
 are presented at the 90\% confidence level.
 
  In addition we  made an attempt to search for their invisible and weak decays via $J/\psi\rightarrow\phi\eta$ and $J/\psi\rightarrow\phi\eta$~\cite{bes3_inv,bes3_weak}.
 These two-body decays provide a very simple event topology, in which the $\phi$ meson can be reconstructed easily and cleanly with its dominant decays
 of $\phi\rightarrow K^+K^-$ .  Since the $\phi$ and $\eta (\eta^\prime)$ are given strong boosts in the $J/\psi$ decay,  
 the invisible decays of the $\eta$  and $\eta^\prime$ were investigated with the mass spectra recoiling against $\phi$.  We find no signal above background for the invisible decays of $\eta$ and $\eta^\prime$. To reduce the systematic uncertainty, the upper limits of the ratios, 
 $\frac{B(\eta\rightarrow invisible)}{B(\eta\rightarrow\gamma\gamma)}<2.6\times 10^{-4}$ and 
  $\frac{B(\eta^\prime\rightarrow invisible)}{B(\eta^\prime\rightarrow\gamma\gamma)}<2.4\times 10^{-2}$, were obtained first at the 90\% confidence level. Then, using the branching fractions of $\eta(\eta^\prime)\rightarrow\gamma\gamma$, the branching fraction upper limits at the 90\% confidence level were determined to be $B(\eta\rightarrow invisible)<1.0\times 10^{-4}$ and $B(\eta^\prime\rightarrow invisible)<5.3\times 10^{-4}$. For the first time  a search for
 the semileptonic weak decay modes $\eta(\eta^\prime)\rightarrow\pi^+e^-\bar{\nu_e}$ was performed and no signal was observed.  At the 90\% confidence level, the semileptonic weak rates were given to be $B(\eta\rightarrow\pi^+e^-\bar{\nu_e}+c.c.)<1.7\times 10^{-4}$ and $B(\eta^\prime\rightarrow\pi^+e^-\bar{\nu_e}+c.c.)<2.2\times 10^{-4}$.

Based on the 225.3 million $J/\psi$ events,  we present the recent results on $\eta$ and $\eta^\prime$ decays in this talk.  To precisely test the fundamental symmetries and theoretical predictions,    the larger statistics of $\eta(\eta^\prime)$ decays are strongly needed. In 2012 the BESIII  detector collected about 1 billion $J/\psi$ events,  four times larger than the sample taken in 2009, which allows us to update the study of $\eta^\prime$, including the Dalitz plot analysis, the search for new decays, as well as the test to the fundamental symmetries. We believe that more interesting results will be coming soon in the near future.

\newpage

\subsection{Hadron Physics Studies at KLOE/KLOE-2}
\addtocontents{toc}{\hspace{2cm}{\sl S.~Giovannella }\par}

\vspace{5mm}

 S.~Giovannella 

\vspace{5mm}

\noindent
Laboratori Nazionali di Frascati dell'INFN,
via Enrico Fermi 40, Frascati, Italy\\
\vspace{5mm}
on behalf of the KLOE-2 Collaboration
\vspace{5mm}

The KLOE experiment at the Frascati $\phi$-factory DA${\Phi}$NE 
collected 2.5 fb$^{-1}$ at the $\phi$ meson peak and about 240 
pb$^{-1}$ below the $\phi$ resonance ($\sqrt{s}=1$ GeV), providing 
large samples of light mesons.
The KLOE-2 detector has been upgraded with small angle tagging devices 
to detect electrons or positrons in $e^+e^-\to e^+e^-X$ events and with 
an inner tracker and small angle calorimeters in the interaction region 
to increase the acceptance both for charged particles and photons. 
A new data taking is planned in years 2014-2015, aiming to collect 
5 fb$^{-1}$. A detailed description of the experimental physics program 
can be found in Ref.~\cite{KLOE2}.

The $\eta\to\pi^+\pi^-\gamma$ decay dynamics has been studied to search
for a possible contribution from chiral anomaly, a higher term of the 
ChPT Lagrangian describing the direct coupling of three pseudoscalar 
mesons with the photon \cite{Benayoun}.
The analysis has been performed using 558 pb$^{-1}$, where about 
$25 \times 10^{6}\ \eta$'s are produced together with a monochromatic 
recoil photon ($E_{\gamma_\phi} = 363$ MeV) through the radiative decay 
$\phi\to\eta\gamma$. 
The process $\eta\to\pi^+\pi^-\pi^0$, with similar event topology and 
negligible background contamination, has been used as normalization 
sample. The ratio of the partial decay widths \cite{KLOEppg},
$ \frac{\Gamma(\eta\to\pi^+\pi^-\gamma)}{\Gamma(\eta\to\pi^+\pi^-\pi^0)} = 
  0.1856\pm 0.0005_{\rm stat}\pm 0.0028_{\rm syst}$,
%
points for a sizable contribution of the direct term to the total width.
The $M_{\pi^+ \pi^-}$ dependence has been parametrized with the model 
independent approach of Ref.~\cite{Kupsc}.

The $\eta\to\pi^+\pi^-\pi^0$ process is an isospin violating decay, 
sensitive to light quark mass difference \cite{Leutw}. 
Dalitz plot analysis, based on 450 pb$^{-1}$, have been performed 
at KLOE in 2008 \cite{KLOE3pi} and have been used in dispersive 
analysis to extract the quark mass ratio \cite{Disp1, Disp2}.
A new high statistics Dalitz plot analysis is in progress with an 
independent and larger (1.7 fb$^{-1}$) data set, using a new analysis 
scheme and improved Monte Carlo (MC) simulation. Preliminary fit 
results, reported in Tab.~\ref{Tab:eta3pi} 
are in agreement with previous KLOE measurement. Evaluation of 
systematics is in progress.
\begin{table}[h]
  \centering
  \caption{Fit results for $\eta\to\pi^+\pi^-\pi^0$ Dalitz plot analysis.}
  \label{Tab:eta3pi}       
  \begin{tabular}{lcccc}
    \hline
               & $a$ & $b$ & $d$ & $f$ \\\hline
    \footnotesize KLOE08     & \footnotesize $-1.090\pm0.005^{+0.008}_{-0.019}$ &
    \footnotesize $0.124\pm0.006\pm0.010$ & \footnotesize $0.057\pm0.006^{+0.007}_{-0.016}$ &
    \footnotesize $0.14\pm0.01\pm0.02$\\
    \footnotesize KLOE prel. & \footnotesize $-1.104\pm0.003$ & \footnotesize $0.144\pm 0.003$ & \footnotesize $0.073\pm0.003$ &
    \footnotesize $0.155\pm0.006$ \\\hline
  \end{tabular}
\end{table}

Pseudoscalar production associated to internal conversion of the 
photon into a lepton pair allows the measurement of the form 
factor $F_P(q_1^2=M_\phi^2\,,q_2^2>0)$ in the kinematical region of 
interest for the VMD model. Detailed study of such decays has been 
performed using 1.7 fb$^{-1}$ of data, both for $\phi\to\eta e^+e^-$
and $\phi\to\pi^0 e^+e^-$ processes.
About 30,000 $\phi\to\eta e^+e^-$, $\eta\to\pi^0\pi^0\pi^0$ candidates 
are present in the analyzed data set, with a residual background 
contamination below 3\%, providing a preliminary measurement of the 
branching fraction: 
${\rm BR}(\phi\to\eta e^+e^-) = 
(1.131 \pm 0.032_{\rm stat+norm}\, ^{+0.011}_{-0.006}\, _{\rm syst} ) 
\times 10^{-4}$.
The resulting electron-positron invariant mass shape, $M_{ee}$, has 
been fitted using the decay parametrization from Ref.~\cite{Landsberg85}.
The preliminary value obtained for the slope of the transition form 
factor in the whole KLOE data set is: $b_{\phi\eta} = 
(1.17 \pm 0.11_{\rm stat}\, ^{+0.09}_{-0.08}\, _{\rm syst} )\ {\rm GeV}^{-2}$,
in agreement with VMD predictions.
For the decay $\phi\to\pi^0 e^+e^-$ no data are available on transition 
form factor. Dedicated analysis cuts strongly reduce the main background 
component of Bhabha scattering events to $\sim 20\%$, which still dominates 
for $M_{ee}>300$ MeV, while a sample of $\sim 9000$ signal candidates is 
obtained. Studies are in progress to refine the evaluation of background 
contamination and of analysis efficiencies.

Data collected at $\sqrt{s} = 1$ GeV have been used to study hadron 
production in $\gamma\gamma$ interactions, providing the most precise 
measurement of the $\Gamma(\eta\to\gamma\gamma)$ partial width from 
the measurement of the $e^+e^-\to e^+e^-\eta$ cross section, using both 
neutral and charged $\eta\to\pi\pi\pi$ decay channels \cite{KLOEggeta}. 
The main background is due to resonant $\phi\to\eta\gamma$ events, with 
an undetected recoil photon. After reducing background components with 
specific kinematical cuts, signal events are extracted by fitting with 
the expected Monte Carlo components the two-dimensional plot 
$M_{\rm miss}^2$--$p_{L/T}$, where $M_{\rm miss}^2$ is the squared missing 
mass and $p_{L/T}$ is the $\eta$ longitudinal/transverse momentum in the
$\pi^0\pi^0\pi^0/\pi^+\pi^-\pi^0$ decay. Combining the two measurements, 
the extracted value for the production cross section is:
$
\sigma(e^+e^-\to e^+e^-\eta) = (32.7 \pm 1.3_{\rm stat} \pm 0.7_{\rm syst})\ {\rm pb}
$
This value is used to extract the most precise measurement of the 
$\eta\to\gamma\gamma$ partial width:
%
$
  \Gamma(\eta\to\gamma\gamma) = 
  (520 \pm 20 _{\rm stat}\pm 13_{\rm syst})\ {\rm eV}$.

\newpage

\subsection{Dispersion theory to connect  $\eta\to \pi \pi \gamma$ to $\eta \to \gamma^* \gamma$}
\addtocontents{toc}{\hspace{2cm}{\sl C.~Hanhart}\par}

\vspace{5mm}

C.~Hanhart

\vspace{5mm}
\noindent
IKP and IAS, Forschungszentrum J\"ulich, Germany\\
\vspace{5mm}

A dispersion integral is derived that connects data on $\eta\to \pi^+\pi^-\gamma$
to the $\eta\to \gamma\gamma^*$ transition form factor~\cite{eta2gammagamma}. 
It is demonstrated that both reactions are controlled by two scales: a universal one
driven by the $\pi\pi$-final state interactions (and of the order of the lightest vector
meson mass) and one that is reaction specific~\cite{stollenwerk}. 
A detailed
analysis of the uncertainties is provided. 
We find for the slope of the  $\eta$ transition form factor at the origin  
 $b_\eta = \left(2.05 \ \mbox{}^{+0.22}_{-0.10} \right)\mbox{GeV}^{-2}$.    
Using an additional, plausible assumption, one finds for the corresponding slope of the
$\eta'$ transition form factor,  $b_{\eta'} = \left(1.58\   \mbox{}^{+0.18}_{-0.13} \right)\mbox{GeV}^{-2}$.
Both values are
consistent with all recent
data, but differ from some previous theoretical analyses.
We regard this study, that provides a systematic improvement compared to the
vector meson dominance approach backed by a sound theoretical method, as an important step towards a better quantitative control 
of the hadronic light-by-light scattering contribution to the muon anomalous magnetic moments~\cite{g22}.

\newpage

\subsection{Dispersion theory and chiral dynamics:\\ from light- to heavy-meson decays}
\addtocontents{toc}{\hspace{2cm}{\sl B.~Kubis}\par}

\vspace{5mm}

B.~Kubis

\vspace{5mm}

\noindent
Helmholtz-Institut f\"ur Strahlen- und Kernphysik (Theorie) and\\ Bethe Center for Theoretical Physics,
Universit\"at Bonn, Germany

\vspace{5mm}

Dalitz plot studies of weak three-body decays of mesons with open heavy flavor
(both $D$ and $B$) may play a key role in future precision investigations of CP violation, within and beyond the Standard Model.
This is due to their much richer kinematic freedom compared to the
(effective) two-body final states predominantly used to
study CP violation at the $B$ factories: the resonance-rich environment of multi-meson
final states may help to enlarge small CP phases in parts of the Dalitz plot~\cite{Hadron2011}.
Traditionally, Dalitz plots have been analyzed experimentally in terms of isobar models:
pairwise interaction between final-state particles, approximated in terms of Breit--Wigner resonances
plus background terms.  However, there is no way to separate resonant from non-resonant contributions 
in a model-independent way; some partial waves, most notably the pion--pion and pion--kaon S-waves 
(of isospin $I=0$ and $I=1/2$, respectively), cannot be modeled in terms of Breit--Wigner functions at all; 
and finally, three-body interactions can modify the isobar picture significantly.

Dispersion relations represent a model-independent method to describe final-state interactions, based on 
input for (re)scattering phase shifts.  If two strongly interacting particles are produced from a point source,
the corresponding form factors can be described in terms of Omn\`es representations; see e.g.\
Ref.~\cite{HanhartFF} for recent work on the pion vector, and Ref.~\cite{Daub} (as well as references therein) 
for the pion scalar form factor.
For three hadrons in the final state, the Khuri--Treiman formalism~\cite{KhuriTreiman} (applied in the formulation
of Ref.~\cite{AnisovichLeutwyler}) allows to write down Omn\`es-like solutions including inhomogeneities,
which are given by partial-wave-projected crossed-channel amplitudes.  
Such a system has been studied for the three-pion decays of the lightest isoscalar vector mesons $\omega$ and $\phi$~\cite{V3pi},
which has been shown to describe the $\phi\to3\pi$ Dalitz plot perfectly.  
The phenomenological contact interactions required in the experimental analysis~\cite{KLOE:phi}, which
necessarily violate unitarity, thereby seem to 
emulate the non-trivial three-body rescattering effects not otherwise included.
Very similar sets of equations can also be used to analyze the anomalous process $\gamma\pi\to\pi\pi$~\cite{g3pi}.

A further application of dispersion theory concerns the vector-meson transition form factors 
as measured in $\omega/\phi\to\pi^0\ell^+\ell^-$~\cite{omegaTFF}: they only require the corresponding three-pion decay amplitudes
and the pion vector form factor as input.  The comparison to experimental data for $\omega\to\pi^0\mu^+\mu^-$ obtained
from heavy-ion reactions~\cite{NA60} however remains problematic.

An ongoing extension of dispersive Dalitz plot analyses concerns the decay $D^+\to\pi^+\pi^+K^-$, with a richer structure
of pion--pion and pion--kaon partial waves, dynamical coupling to the $\pi^+\pi^0\bar K^0$ final state~\cite{BESIII:DKspi0pi},
and a larger number of subtraction constants to be fixed.  
Preliminary fits to data~\cite{CLEO:DKpipi} in the kinematic region where elastic unitarity in $\pi K$ scattering 
should still hold to good accuracy suggest a similar improvement through three-body rescattering as for the $\phi\to3\pi$
Dalitz plot~\cite{Franz:D}.  Whether or not such three-body final states can be used to actually \emph{learn} something
about $\pi K$ scattering phases in a model-independent way remains to be investigated~\cite{E791:DKpipi,FOCUS:DKpipi}.

The latter may be more straightforward for semileptonic decays, such as $D\to \pi K \ell\nu$, without a third
strongly-interacting particle in the final state.  However, even in such a case, left-hand singularities may be important, 
as has been demonstrated for the similar process $B\to\pi\pi\ell\nu$~\cite{Bl4},
an exclusive decay channel that potentially allows for an extraction of the CKM matrix element $|V_{ub}|$.
$B^*$-pole terms dominate the amplitude at leading order in heavy-meson chiral perturbation theory~\cite{Burdman+Wise}
in the kinematic region of two very soft pions; dispersion theory allows to vastly extend the kinematic range of applicability,
and to control the shape of the partial waves, including S-wave background to the presumed $\rho$ dominance.

\newpage

\subsection{Interactions of light with light mesons}
\addtocontents{toc}{\hspace{2cm}{\sl  S.~Leupold}\par}

\vspace{5mm}

S.~Leupold

\vspace{5mm}

\noindent
Department of Physics and Astronomy, Uppsala University, Sweden

\vspace{5mm}

Electromagnetic probes are a good way to explore the intrinsic structure of hadrons. A second reason why the interactions 
between light mesons and photons are interesting comes from the present disagreement between the experimental value of the 
gyromagnetic ratio of the muon and its standard-model prediction 
(see, e.g., \cite{Jegerlehner:2009ry,Czerwinski:2012ry} and references therein). 
The hadronic contributions to this gyromagnetic ratio
constitute the largest uncertainty in the standard-model prediction. These contributions can be split into the hadronic vacuum
polarization and the light-by-light scattering contribution. The former is directly related to a measurable quantity via 
dispersion theory. At present the latter requires hadronic-theory input. 
This calls for high-precision experiments and for a hadronic 
theory where the uncertainties can be reliably estimated. Concerning hadronic calculations in the resonance region we have not
reached this aim yet. But steps are undertaken in this direction exploring different techniques and concepts.

One approach is based on a chiral Lagrangian for pseudoscalar and vector mesons complemented by rules how to assign specific 
levels of importance to various Feynman diagrams. (On purpose the phrase ``power counting'' is avoided until the impact of 
higher-order contributions has been studied systematically. This is presently under investigation.) The approach is well
documented in the 
literature \cite{Lutz:2008km,Leupold:2008bp,Terschluesen:2010ik,Danilkin:2011fz,Terschlusen:2012xw,Danilkin:2012ua,matthias}. 
Relations
to light-by-light scattering are manifold: Reactions of two photons to two pseudoscalar mesons are studied in 
\cite{Danilkin:2012ua}. Due to the intimate relation between photons and (neutral) vector mesons (same quantum numbers) 
electromagnetic transition form factors between vector and pseudoscalar 
mesons \cite{Terschluesen:2010ik,Terschlusen:2012xw} constitute particular kinematical situations of the coupling of a 
single pseudoscalar meson to two virtual photons.

\begin{figure}[h]
  \centering
  \includegraphics[keepaspectratio,width=0.39\textwidth]{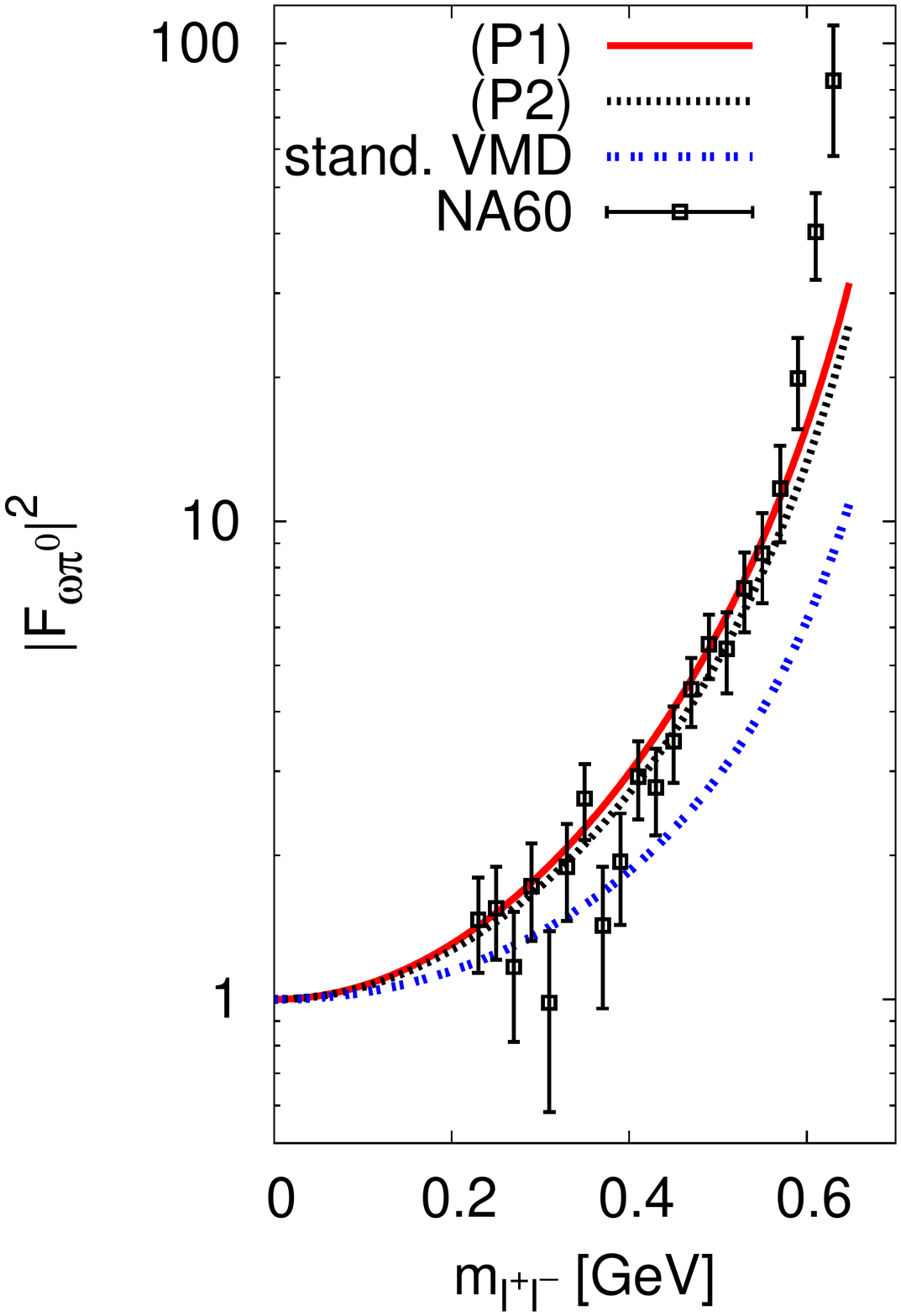} \hfill
  \includegraphics[keepaspectratio,width=0.59\textwidth]{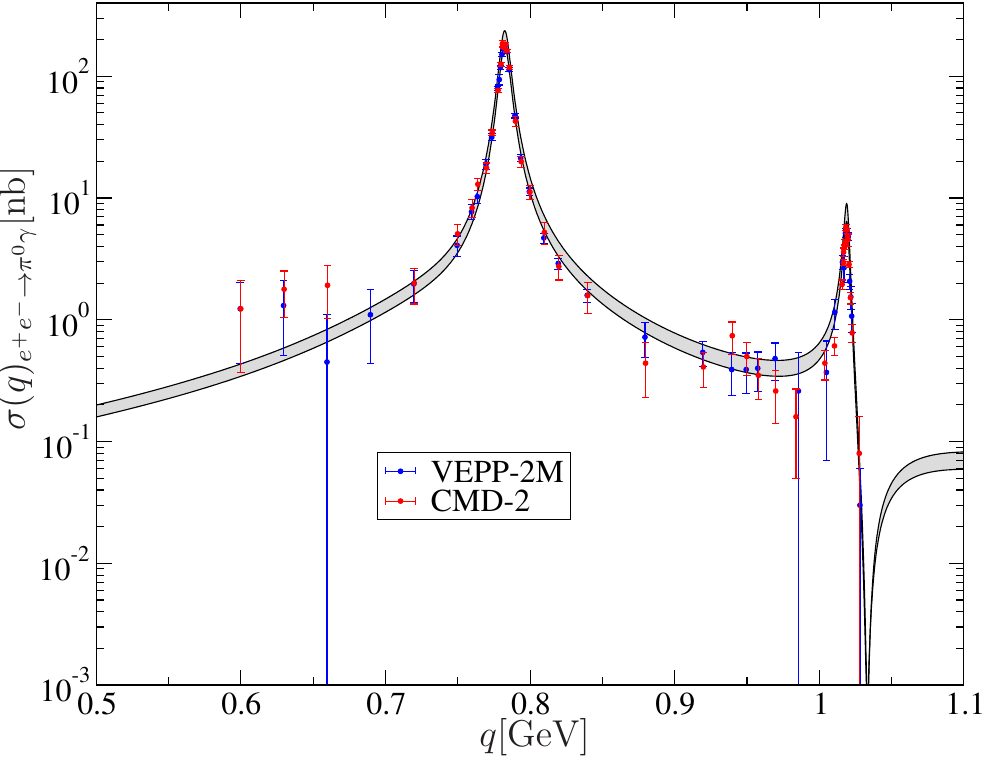}
  \caption{{\it Left-hand side:} The $\omega$ transition form factor from the Lagrangian approach (full red line) as 
    compared to data
    from the NA60 experiment \cite{Arnaldi:2009aa} and to the result from standard vector-meson dominance (blue dashed line). See
    \cite{Terschluesen:2010ik} for details.
    {\it Right-hand side:} The cross section for $e^+ e^- \to \gamma \pi^0$ (pion transition form factor) 
    from the dispersive approach \cite{wip} as compared to data from Novosibirsk \cite{Achasov:2003ed,Akhmetshin:2004gw}.}
  \label{fig:fig}
\end{figure}

To highlight one result of the Lagrangian approach the electromagnetic transition form factor for $\omega$ to $\pi^0$ 
is depicted on the left-hand side of figure \ref{fig:fig}.
Obviously the overall description is good and much better than the traditional standard vector-meson dominance model. However,
there is a clear mismatch between the NA60 dimuon data and calculations for the high-mass region close to the phase-space 
limit of the reaction
$\omega \to \pi^0 \mu^+ \mu^-$. An independent confirmation of 
the NA60 results from a more exclusive experiment, e.g.\ in the decay reaction $\omega \to \pi^0 e^+ e^-$, would be 
extremely welcome.

A second approach to meson transition form factors is based on dispersion theory and excellent data for pion phase shifts and the
(direct) pion vector form factor \cite{Niecknig:2012sj,Schneider:2012ez,Hoferichter:2012pm,Hanhart:2013vba,proc2}. 
For the calculation of the pion-to-photon transition form factor one analyzes the possible hadronic
inelasticities. For the reaction $e^+ e^- \to \gamma \pi^0$ up to about 1 GeV the relevant intermediate hadronic states are
two pions (isospin 1) or three pions (isospin 0). Thus for the isospin-1 case the imaginary part of the scattering amplitude is
given by the consecutive reactions $e^+ e^- \to \pi^+ \pi^-$ and $\pi^+ \pi^- \to \gamma \pi^0$. The former is just the 
well-known pion vector form factor, the latter has been calculated in the dispersive approach in \cite{Hoferichter:2012pm}.
Note that the previously discussed omega transition form factor has also been calculated in the dispersive 
approach \cite{Schneider:2012ez}. A fully dispersive treatment of the case where the dielectron has isospin 0 in the 
reaction $e^+ e^- \to \gamma \pi^0$ would require a proper handling of the amplitudes $e^+ e^- \to \pi^+ \pi^-\pi^0$ 
and $\pi^+ \pi^- \pi^0 \to \gamma \pi^0$. This is technically beyond the scope of present works and also lacks the necessary
differential data as input. However, the isospin-0 case is dominated at low energies by the narrow resonances 
$\omega$ and $\phi$. Therefore a dispersively improved Breit-Wigner approach is pursued here. Parameters (peak positions,
peak heights and polynomial background terms) are chosen such that the reaction 
$e^+ e^- \to \pi^+ \pi^-\pi^0$ is properly described. Using an unsubtracted dispersion relation, one obtains the pion
transition form factor displayed in figure \ref{fig:fig}, right-hand side. In \cite{Schneider:2012ez} it
has been shown that the decay widths $\omega,\phi \to \gamma \pi^0$ as obtained by the same technique are precise on a 
ten percent level. The 
corresponding uncertainty has been added to the isospin-0 part of the pion transition form factor resulting in the gray band
shown in figure \ref{fig:fig}, right-hand side. Obviously a very decent description of the pion transition form factor can be 
obtained in this way \cite{wip}. It opens the way for the calculation of the spacelike part of the pion transition form factor
and eventually for the corresponding double-virtual form factor, which in turn enters as one important contribution into 
the light-by-light scattering amplitude. 

Several present high-statistics experiments allow for detailed studies of rare meson decays in 
the meson-mass range of 1 GeV. 
In particular the $\eta'$ meson obtains and deserves a lot of attention since its properties are intimately related to the 
chiral anomaly. The vector mesons in this mass range seem to be much less appealing since they are ordinary quark-antiquark 
states and their main properties are not dominated in a fancy way by any broken or unbroken symmetry. 
However, electromagnetic probes of hadrons cannot be understood (in particular at a high-precision level) without 
understanding the vector mesons. This remark even applies to the $\eta'$ decays involving photons from which one would 
like to learn something about the chiral anomaly. This strongly suggests to study not only $\eta'$ decays but with the 
same dedication also rare decays of vector mesons, in particular also of omega mesons. For instance improved experimental 
differential data for the poorly known reactions $\omega/\phi \to \pi^0\,e^+ e^-$ and $\omega \to \gamma+2\pi$ would be highly 
desirable. Among other aspects they would help to sharpen the theory tools outlined above. Both mentioned reactions have a 
clear connection to light-by-light scattering: The decay $\omega/\phi \to \pi^0\,e^+e^-$ is intimately related to one 
particular kinematical region of the pion transition form factor where one invariant mass is fixed to the vector-meson mass. 
The decay $\omega \to \gamma+2\pi$ provides the lowest-energetic inelasticity for $\omega \to 3\gamma$ which in turn is 
directly related to light-by-light scattering with three photons onshell and one with the invariant mass of the $\omega$ meson.

\newpage

\subsection{Chiral dynamics with vector mesons}
\addtocontents{toc}{\hspace{2cm}{\sl  M.F.M.~Lutz}\par}

\vspace{5mm}

M.F.M.~Lutz

\vspace{5mm}

\noindent
GSI Helmholtzzentrum f\"ur Schwerionenforschung GmbH\\ Planck Str. 1, 64291 Darmstadt, Germany\\

\vspace{5mm}

The light vector mesons play a crucial role in the hadrogenesis
conjecture \cite{Lutz:2001dr,Lutz:2003fm,Lutz:2004dz,Lutz:2005ip,Lutz:2007sk,Lutz:2008km,Terschluesen:2012}.
Together with the Goldstone bosons they are identified to be the ``quasi-fundamental'' hadronic degrees of freedom that are expected to generate the meson spectrum. For instance it was shown that the leading chiral interaction of Goldstone bosons with the light vector mesons generates an axial-vector meson spectrum that
is quite close to the empirical spectrum \cite{Lutz:2003fm}. 

Though it is well known how to incorporate more massive degrees of freedom into the chiral Lagrangian, it is a challenge 
how to organize systematic applications. The key issue is the identification of an optimal set of degrees of freedom in combination with the construction of power counting rules. A novel counting scheme for the chiral Lagrangian which includes the nonet of light vector mesons in the tensor field representation was explored in \cite{Lutz:2008km,Terschluesen:2012}. It is based on the hadrogenesis conjecture and large-$N_c$ considerations \cite{Lutz:2001dr,Lutz:2003fm,Lutz:2004dz,Lutz:2005ip,Lutz:2007sk}. The counting scheme would be a consequence of an additional mass gap of QCD in the chiral limit, that may arise if the
number of colors increases. Whether it leads to a fully systematic effective field theory is an open issue.

The leading-order hadrogenesis Lagrangian as constructed in \cite{Lutz:2008km,Terschluesen:2012} was tested in various applications so far. Most of the low-energy parameters can be estimated by hadronic and electromagnetic properties of the vector mesons evaluated at tree-level \cite{Lutz:2008km,Terschluesen:2012,Leupold:2009,Terschluesen:2010}.
Applications to coupled-channel systems  \cite{Danilkin:2011,Danilkin:2012} are based on a novel unitarization scheme 
that is justified in the presence of long and short-range forces \cite{Gasparyan:2010xz,Danilkin:2010xd,Gasparyan:2011yw,Danilkin:2011,Danilkin:2012,Gasparyan:2012km}. A first systematic analysis of pion-pion and  pion-kaon scattering can be found in \cite{Danilkin:2011}. Given the fact that at leading order there is no free parameter a remarkable reproduction of the empirical phases shifts was obtained. Like in previous coupled-channel studies of such systems various scalar resonances are generated dynamically. As a further application photon fusion reactions were considered in \cite{Danilkin:2012}. In this case a few low-energy parameters were adjusted to the data set. In our scheme already at leading order the vector-meson exchange processes play an important role. This is contrasted by the standard $\chi$PT approach where the leading order interactions are not affected by vector-meson exchange processes \cite{Ecker:1988te}. In this case the subleading counter terms my be estimated by a saturation ansatz in terms of  vector-meson exchange processes \cite{Ecker:1988te}. 

At higher energies further resonances come into play. In principle,
also for instance the tensor resonances $f_2(1270)$ and $a_2(1320)$ are
expected to be naturally generated within our approach from
vector-vector interactions \cite{Lutz:2007sk}. However, a significant application of the 
hadrogenesis Lagrangian to the scattering of two vector mesons is quite a challenge. So far no realistic computations based on the chiral Lagrangian have been performed. 
The basis for the systematic inclusion of pairs of vector mesons as coupled-channel states has been laid
out recently in \cite{Lutz:2012}. In a first step it is necessary to identify partial-wave amplitudes that 
have convenient analytic properties \cite{Gasparyan:2010xz,Stoica:2011,Lutz:2012} as to be used in the unitarization scheme \cite{Gasparyan:2010xz,Danilkin:2010xd,Gasparyan:2011yw,Danilkin:2011,Danilkin:2012,Gasparyan:2012km}. 
Once intermediate states with two vector mesons are considered there are almost always long-range forces 
impled by t- or u-channel exchange processes that lead to non-trivial left-hand branch points in the partial-wave
amplitudes. The positions of left- and right-hand branch cuts almost always overlap. 

There is a subtle limitation of algebraic or separable approaches, if applied to such a coupled-channel situation. The partial-wave scattering amplitudes have necessarily unphysical left-hand branch points \cite{Lutz:2003fm,Danilkin:2011}. This holds at any finite truncation unless the K-matrix ansatz, which is at odds with micro causality, is imposed. Though such unphysical left-hand branch violate the micro-causality condition they do not necessarily always lead to numerically significant effects in the physical region. If all considered left-hand branch points are below the smallest considered threshold, i.e. the left- and right hand branch cuts do not overlap, the presence of unphysical branch points are not really problematic. However, once a left-hand branch point of a heavy channel like the two-vector meson channel is located right to the threshold of a lighter channel an algebraic approach can no longer be justified.

\newpage

\subsection{$\eta$ Transition Form Factors from Rational Approximants}
\addtocontents{toc}{\hspace{2cm}{\sl  P.~Masjuan}\par}

\vspace{5mm}

 P.~Masjuan

\vspace{5mm}

\noindent
PRISMA Cluster of Excellence, Institut f\"ur Kernphysik, Johannes Gutenberg-Universit\"at Mainz, D-55099 Mainz, Germany\\

\vspace{5mm}

The pseudoscalar transition form factor (TFF) describes the effect of the strong interaction on the $\gamma^*\gamma^* - P$ transition, where $P=\pi^0, \eta, \eta^\prime \dots$, and is represented by a function $F_{P\gamma^*\gamma^*}(q_1^2,q_2^2)$ of the photon virtualities $q_1^2$, and $q_2^2$.

From the experimental point of view, one can study such TFF from both space-like and time-like energy regimes. The time-like TFF can be accessed from a single Dalitz decay $P \to l^+l^- \gamma$ process which contains an off-shell photon with the momentum transfer $q_1^2$ and defines a $F_{P\gamma^*\gamma}(q_1^2,0)$ covering the $4m_l^2<q^2<m_P^2$ region.  The space-like TFF can be accessed in $\ee$ colliders by the two-photon-fusion reaction $e^+e^-\to e^+e^-P$. The common practice is to extract the TFF when one of the outgoing leptons is tagged and the other is not, that is, the single-tag method.
The tagged lepton emits a highly off-shell photon with the momentum transfer $q_1^2\equiv -Q^2$ and is detected,
while the other, untagged, is scattered at a small angle and its momentum transfer $q_2^2$ is near zero, i.e., 
$F_{P\gamma^*\gamma}(Q^2)\equiv F_{P\gamma^*\gamma^*}(-Q^2,0)$.

Theoretically, the limits $Q^2=0$ and $Q^2\rightarrow\infty$ are well known in terms of the axial anomaly in the chiral limit of QCD \cite{Adler:1969gk} and pQCD \cite{Lepage:1980fj}, respectively. The TFF is then calculated as a convolution of a perturbative hard-scattering amplitude and a gauge-invariant meson distribution amplitude (DA)~\cite{Mueller:1994cn} which incorporates the nonperturbative dynamics of the QCD bound-state~\cite{Lepage:1980fj}. Some model needs to be used either for the DA or the TFF itself. The discrepancy among different approaches reflects the model-dependency of that procedure. A different procedure might be, then, desirable.

We propose~\cite{EscribanoMasjuan}  to use a sequence of rational approximants called Pad\'e approximants (PA)~\cite{Baker} constructed from the Taylor expansion of the $F_{\eta\gamma^*\gamma}(Q^2)$ to fit the space- and time-like experimental data, Refs.~\cite{SL} and~\cite{TL} resp., and obtain, in such a way, the derivatives of the $F_{\eta\gamma^*\gamma}(Q^2)$ at the origin of energies in a simple, systematic and model-independent way~\cite{Masjuan:2012wy}. Including the decays of the $\eta^{(')}\rightarrow \gamma \gamma$ in our set of data, we can systematically predict the slope and the curvature of both $\eta^{(')}$-TFFs. The low-energy parameters obtain with this method can be used to constrain the hadronic models used to account for the light-by-light scattering contribution part of the anomalous magnetic moment of the muon~\cite{EscribanoMasjuan,g2}, rare $\eta$ decays and continuum cross section determinations in the charmonium region~\cite{EscribanoMasjuan}. Reference~\cite{EscribanoMasjuan} also provides with parameterizations for such form factors valid for the whole space-like energy range. Notice, however, that even though the procedure followed here is based on model-independent methods, the PA fit does not provide with an extraction of the resonance pole exchanged in the process since PA cannot be analytically continued into the complex plain where poles are supposed to lie~\cite{Masjuan:2013jha}. The same comment applies for interpreting the outcome of a fit with a Vector Meson Dominance model (its pole parameter) as the vector meson mass participating in the process~\cite{Masjuan:2007ay}.

The physical $\eta$ and $\eta'$ mesons are an admixture of the $SU(3)$ Lagrangian eignestates~\cite{Leutwyler:1997yr}. Deriving the parameters governing the mixing is a challenging task. Usually, these are determined through the use of $\eta(')\rightarrow \gamma\gamma$ decays as well as vector radiative decays into $\eta(')$ together with $\Gamma(J/\Psi \to \eta^\prime \gamma)/\Gamma(J/\Psi \to \eta \gamma)$~\cite{Leutwyler:1997yr}. However, since pQCD predicts 
that the asymptotic limit of the TFF for the $\eta(')$ is essentially given in terms of these mixing 
parameters~\cite{Feldmann:1998yc}, we use our TFF parametrization to estimate the asymptotic limit and further constrain the mixing parameters with compatible results compared to standard (but more sophisticated) determinations.

\newpage

\subsection{$\eta-\pi$ isospin violating form factors}
\addtocontents{toc}{\hspace{2cm}{\sl B.~Moussallam}\par}

\vspace{5mm}
B.~Moussallam

\vspace{5mm}

\noindent
Groupe de physique théorique, IPN, Université Paris-Sud-11, Orsay, France\\
\vspace{5mm}

The fact that  isospin symmetry appears to be nearly exact in
nature is linked to the peculiar mass pattern   of the three
lightest quarks in QCD: $m_d-m_u << m_s$  and $m_s << 1 $ GeV. One of
the goals of the chiral effective low-energy theory of QCD is to
arrive at a consistent and precise determination of ratios of the
light quark masses  based on experimental measurements involving 
light  mesons.  At present, the determination of isospin
breaking quark mass ratios like
$1/Q^2=(m_u^2-m_d^2)/(m_s^2-m_{ud}^2)$ from different observables
lead to differences as large as 20\%. This has triggered efforts, on
the experimental side,  for performing improved  measurements of the
$\eta\to 3\pi$
amplitude\cite{Adolph:2008vn,Prakhov:2008ff1,Unverzagt:2008ny1,Ambrosinod:2010mj1,Ambrosino:2008ht1}, which is  proportional to
$1/Q^2$ to a high acuracy. On the theory side, it was proposed to
supplement the chiral expansion with dispersive methods in order to
improve the treatment of the final-state
interactions~\cite{Kambor:1995yc,Anisovich:1996tx}, based on the
framework originally proposed by Khuri and
Treiman~\cite{Khuri:1960zz}.  

There has also been progress in measurements of isospin breaking in
$K_{l3}$ form factors (see~\cite{Antonelli10}). We reconsider here the
analogous $\eta_{l3}$ form factors, which are vanishing in the isospin
limit, and discuss their relation with the $\eta\to 3\pi$ amplitude and
with the $K_{l3}$ form factors. While the branching fraction for
$\eta_{l3}$ decays are too small for observation, this is not the case for the
$\tau$ decay mode: $\tau\to \eta\pi \nu$. The related isospin suppressed
$\eta-\pi$ form factors could in principle be measured with some precision
at future $\tau$-charm factories and at Belle-II which was not
possible at past $B$ factories because large nackgrounds. This has
modtivated us to reconsider the evaluation of these isospin violating
form factors. References to previous work on this subject can be found
in ref.~\cite{Nussinov:2008gx}.

\begin{figure}[h]
\bc
\includegraphics[width=0.6\linewidth]{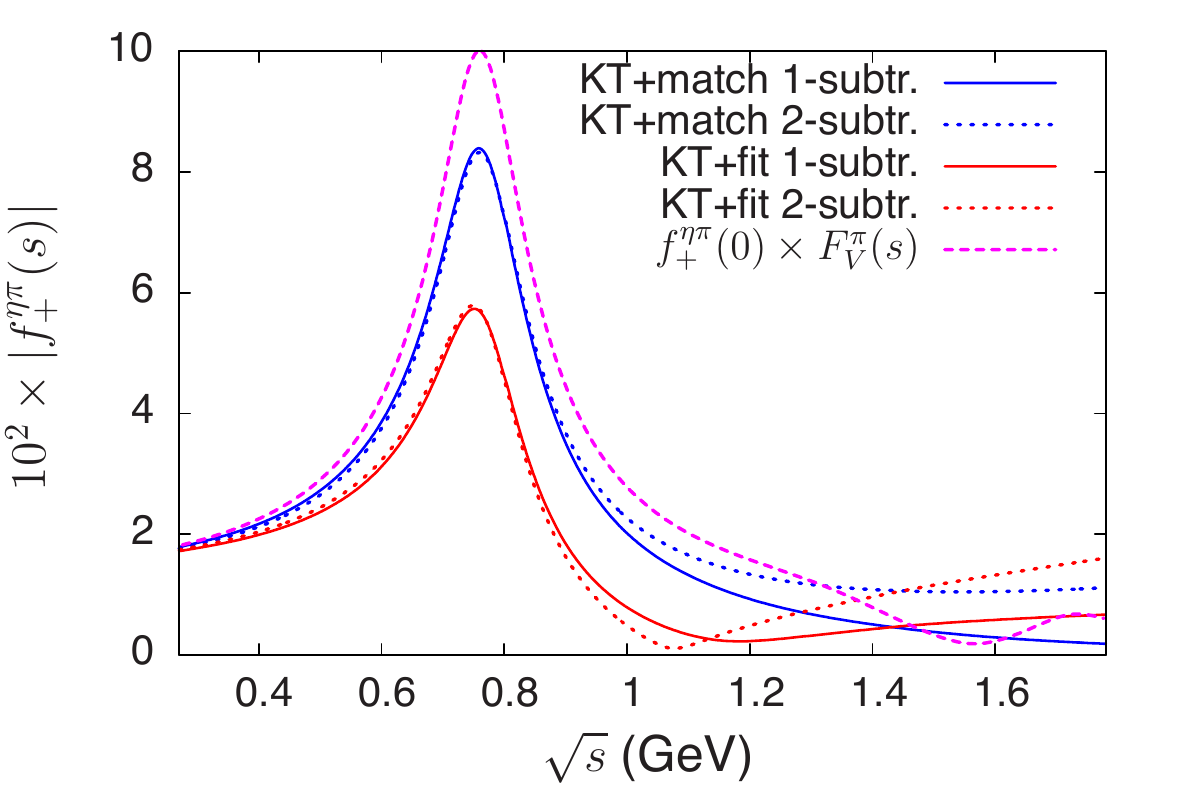}
\caption{\sl Vector $\eta-\pi$ form factor computed from
  eq.~\rf{unitfplus} and different sets of $\eta\to3\pi$ Khuri-Treiman
  solutions and compared to a naive VMD shape. }
\label{fig:fplus}
\ec
\end{figure}

The basic method for the evaluation of the $\eta-\pi$ form factors in
the energy region of the light resonances is to combine ChPT results
with general properties of analyticity and unitarity. From
analyticity, one can write a dispersive representation for the vector
form factor,
\begin{equation}\label{dispfplus}
\fplus(s)=\fplus(0) + s\dotfplus(0)+{s^2\over\pi}\int_{4\mpid}^\infty
ds'\, { \disc[ \fplus(s')]\over (s')^2  (s'-s)}\ .
\end{equation}
One can show (using the method of ref.~\cite{Mandelstam:1960zz}) that
the usual analyticity preperties (in particular, the absence of
anomalous thresholds) holds in the present case in spite of the fact
that the $\eta$ meson is unstable. The values of the form factor and
its derivative at $s=0$, needed in eq.~\ref{dispfplus} may be taken
from the NLO chiral
calculations~\cite{Neufeld:1994eg,Scora:1995sj}. In particular, a
chiral low-energy theorem was derived in ref.~\cite{Neufeld:1994eg}
which enables one to relate $\fplus(0)$ with the $K_{l3}$ form factor
ratio $f_+^{K^+\pi^0}(0)/f_+^{K^0\pi^+}(0)$. Next, one can express
$\disc[ \fplus(s')]$ in eq.~\ref{dispfplus} using unitarity: the
dominating contribution below 1 GeV reads, 
\be\lbl{unitfplus}
 \disc\left[\fplus(s)\right]_{\pi\pi}=- \theta(s-4\mpid) {s-4\mpid\over
    16\pi\, \lambetapi{s} }\, F_V^\pi(s)
\times {1\over2}\int_{-1}^1 dzz\, T^*_{\pi^0\pi^+\to \eta\pi^+}(s,t(z))\ .
\en
It is proportional to the well known pion form factor $F_V^\pi$ and to the
$\eta\to \pi^0\pi^+\pi^-$ amplitude, projected on the $P$-wave. This
amplitude is needed partly in an unphysical region: we have determined
it based on a four-parameter family of numerical solutions of the
Khuri-Treiman equations. These four parameters can be determined
completely from the NLO chiral amplitude by solving a set of four
matching equations. Doing so, one predicts the Dalitz plot parameters
for the charged $\eta$ decay mode to be slightly different from those
measured~\cite{Ambrosino:2008ht}, in particular the Dalitz parameter
$d$ which probes the $(t-u)^2$ dependence, is found to be  larger by
$\simeq 30\%$.  The sensitivity of the form factor shape to the
$\eta$ decay amplitude is illustrated by fig.~\fig{fplus} which
compares the results from an amplitude as predicted by matching and
from an amplitude fitted to experiment.  

A full estimate of the $\eta\pi$ spectral function also requires input for the
scalar form factor $\fzero$. This form factor contains information on
the nature of the scalar resonance $a_0(980)$ via its coupling to the
$\bar{u}d$ scalar operator. We have peformed an estimate based on ChPT
and dispersion relations: in this framework an exotic nature manisfest
itself by the presence of a zero.
\vspace{-0.33cm}

\newpage

\subsection{Effective Field Theories for Vector
    Particles\\ and Constraint Analysis}
\addtocontents{toc}{\hspace{2cm}{\sl A.~Neiser}\par}

\vspace{5mm}
A.~Neiser, S.~Scherer
\vspace{5mm}

\noindent
Institut f\"ur Kernphysik, Johannes Gutenberg-Universit\"at Mainz, Germany\\

\vspace{5mm}

The strong interaction is described by quantum chromodynamics (QCD), a
gauge theory with quarks and gluons as the fundamental particles.
However, experimentally one observes baryons and mesons as bound
states which can be arranged in representations of the flavor symmetry
group SU(3). If one assumes Lorentz invariance and the cluster
decomposition principle, one can formulate an effective theory for the
strong interaction in the low-energy regime by using the most general
Lagrangian consistent with the assumed
symmetries~\cite{Weinberg:1996}. This concept is well-established as
chiral perturbation theory for the quasi-Goldstone bosons, namely,
pions, kaons and etas. In order to extend its applicability to higher
energies, vector mesons such as the rho meson triplet should be
included.

Since effective Lagrangians for vector particles (spin $S=1$, parity
$P=-1$) are constructed with Lorentz four-vectors $V^\mu$ or
anti-symmetric tensors $W^{\mu\nu}=-W^{\nu\mu}$ with four and six
independent fields, respectively, one inevitably introduces more
degrees of freedom than physically realized for a massive spin-one
particle, which are $2S+1=3$. Hence, suitable constraints are needed
to eliminate the unphysical degrees of freedom in order to obtain a
consistent theory already on a classical level. Typically, this leads
to conditions for the numerous coupling constants in the Lagrangians,
which can be helpful in the determination of those a-priori unknown
low-energy constants using experimental data.

In the following, the results of such a constraint analysis are
presented for two examples. It is a short summary of the diploma
thesis by the first author, see chapters 4 to 6 in
Ref.~\cite{Neiser:2011}. The mathematically precise description can be
found in Ref.~\cite{Gitman:1990}.

The first example is an effective theory for eight vector particles,
assuming a \emph{global} SU(3) symmetry. The Lagrangian reads
\begin{equation*}
  \mathcal{L}=-\frac{1}{4}V^a_{\mu\nu}V^{a\mu\nu}+\frac{M^2}{2}V^a_\mu
    V^{a\mu}
  -g^{abc}V^a_{\mu}V^b_{\nu}\pd^\mu V^{c\nu}-h^{abcd}V^a_{\mu}V^b_{\nu}V^{c\mu}V^{d\nu}\,,
\end{equation*}
where $V^a_{\mu\nu}=\pd_\mu V^a_\nu-\pd_\nu V^a_\mu$ and the indices
$a, b, c, d$ range from $1$ to $8$. Owing to the assumed SU(3)
symmetry, the Lagrangian can be parametrized with five real couplings
as follows,
\begin{equation}
  \label{eq:param}
  g^{abc}=\gamma_1f^{abc}+\gamma_2d^{abc}\,,\qquad h^{abcd}=\eta_1\delta^{ac}\delta^{bd}+
  \eta_2\delta^{ab}\delta^{cd}+\eta_3f^{abe}f^{cde}\,,
\end{equation}
where $f_{abc}=\frac{1}{4i}\Tr\bigl([\lambda_a,
\lambda_b]\lambda_c\bigr)$ and
$d_{abc}=\frac{1}{4}\Tr\bigl(\{\lambda_a, \lambda_b\}\lambda_c\bigr)$
with the Gell-Mann matrices $\lambda_a$. If one requires (1) that the
number of constraints reduces the degrees of freedom to the physical
number, and (2) that those constraints are conserved in time on a
classical level, then the following conditions must hold,
\begin{equation}
  \label{eq:constraint}
  \gamma_2=0\qquad\text{and}\qquad\eta_1=-\eta_2\,.
\end{equation}
Next, this result from the constraint analysis can be used in a
subsequent renormalizability analysis. It requires that the infinite
parts resulting from one-loop contributions can be absorbed in the
bare parameters of the Lagrangian, i.\,e. into the vertices at tree
level, which is a necessary but not sufficient condition for a
physically meaningful theory. Since
\begin{equation*}
  h^{1111}\stackrel{\eqref{eq:param}}{=}\eta_1+\eta_2\stackrel{\eqref{eq:constraint}}{=}0\,,
\end{equation*}
the infinite parts of the one-loop contribution to the vertex $V^1V^1V^1V^1$ must
vanish, so
\begin{equation*}
  0=9\Bigl(\gamma_1^2-4\eta_3-\frac{16}{3}\eta_1\Bigr)^2+192\eta_1^2
  \quad\Leftrightarrow\quad
  \eta_1=0,\quad \gamma_1^2=4\eta_3\,.
\end{equation*}
Finally, a massive Yang-Mills theory is obtained with one parameter
$\gamma_1$. This result is similar to Ref.~\cite{Djukanovic:2010},
where the same argument leads from a theory with three vector fields
and a global U(1) symmetry to a massive Yang-Mills theory.

The second example uses anti-symmetric tensor fields
$W^{\mu\nu}=-W^{\nu\mu}$ to describe three massive vector particles.
The Lagrangian reads 
\begin{equation}
  \label{eq:lag-tensor}
  \begin{split}
    \Lag&=-\frac{1}{2}\pd^\mu\W{a}{}{\mu\nu}{}\pd_\rho\W{a}{\rho\nu}{}{}
    +\frac{M_a^2}{4}\W{a}{\mu\nu}{}{}\W{a}{}{\mu\nu}{}
  -g^{abc}\W{a}{}{\mu\nu}{}\W{b}{\mu\lambda}{}{}\W{c}{\nu}{\lambda}{}\\
 &\relphantom{=}{}-h^{abcd}_1\W{a}{\alpha\beta}{}{}\W{b}{\gamma\delta}{}{}
   \W{c}{}{\alpha\beta}{}\W{d}{}{\gamma\delta}{}
  -h^{abcd}_2
  \W{a}{\alpha\beta}{}{}\W{b}{\gamma\delta}{}{}\W{c}{}{\alpha\gamma}{}\W{d}{}{\beta\delta}{}\,,
  \end{split}
\end{equation}
where $M_1=M_2=M$ and identical Lorentz structures resulting from two
Levi-Civita tensors $\epsilon^{\alpha\beta\gamma\delta}$ are omitted.
Assuming U(1) invariance, the couplings can be parametrized using
$1+10=11$ real couplings, 
\begin{align*}
  g^{123}&=g_1\,,\quad h_1^{1111}=h_1^{2222}=d_1\,,\quad h_1^{3333}=d_5\,,\quad
  h_2^{1212}=-4(2d_2+d_7)\,,\\
  h_1^{1122}&=2(d_1-d_2)\,,\quad
  h_2^{1111}=h_2^{2222}=2(d_6-d_1)\,,\quad h_1^{1212}=2d_2\,,\quad h_2^{3333}=2(d_{10}-d_5)\,,
  \\
  h_1^{1133}&=h_1^{2233}=d_3\,,\quad h_2^{1133}=h_2^{2233}=-2(d_3-d_4+d_8+d_9)\,,\\
  h_2^{1313}&=h_2^{2323}=2(d_9-2d_4)\,,\quad h_1^{1313}=h_1^{2323}=d_4\,,\quad
  h_2^{1122}=4(2d_2-d_1+d_6+d_7)\,,
\end{align*}
all other coupling are set to zero without loss of generality. The
constraint analysis yields $g_1=0$, i.\,e. the three-vertex vanishes
completely, and $d_6=d_7=d_8=d_9=d_{10}=0$. Again, the subsequent
renormalizability analysis uses those results and yields
$d_5=d_4=d_3=d_2=d_1=0$. In summary, all interaction terms must vanish
in Eq.~\eqref{eq:lag-tensor}, if one requires U(1) invariance,
self-consistency with respect to constraints, and renormalizability.
This result is completely different from the one using the vector
formalism, and one should investigate tensor models including
interactions with derivatives. However, such an extended analysis
certainly needs computer assistance and smarter implementations, but
it may yield valuable conditions for the numerous coupling constants
in effective Lagrangians.

\newpage

\subsection{Pseudoscalar-vector and vector-vector
interaction and resonances generated}
\addtocontents{toc}{\hspace{2cm}{\sl E.~Oset}\par}

\vspace{5mm}

E.~Oset

\vspace{5mm}

\noindent
 Departamento de F\'iısica Te\'orica and IFIC, Centro Mixto Universidad
de Valencia-CSIC, Institutos de Investigaci\'on de Paterna,
Apartado 22085, 46071 Valencia, Spain\\

\vspace{5mm}

The local hidden gauge approach \cite{hidden1} provides an extension of the chiral Lagrangians, including the interaction of vector mesons among themselves and with pseudoscalars and baryons. It has been used with success to study the interaction of vector mesons in \cite{raquel,gengvec}, where many resonances are dynamically generated as a consequence of this interaction. The picture offers also a good interpretation for the radiative decays of resonances into $\gamma \gamma$ and other decay channels \cite{yama}. In the baryon sector it also gives rise to many baryonic resonances \cite{angelsvec,sourav,javier,review}. 

One of the surprises was the realization that there are two $K_1(1270)$ resonances, separated by about 100 MeV  \cite{roca} and experimental evidence was found in \cite{gengaxial}. 

Very recently we have shown \cite{alba} that an $h_1$ resonance predicted around 1800 MeV in \cite{gengvec} has found experimental support from a BES experiment \cite{besexp}. Similarly we have also shown \cite{zouetal} that one can explain within this picture the decays of $J/\psi$ into $\omega(\phi)$ and the resonances which are made up of two vectors in \cite{gengvec}. Also the decay of $J/\psi$ into a photon and one of these resonances is well described in \cite{hanhart}. The same occurs with the decays of the excited states of $J/\psi$ or of the $\Upsilon$ \cite{dai}.

 The local hidden gauge approach has also allowed us to give a different interpretation of the
peak seen in the threshold of the $\omega \phi$ mass distribution in \cite{oldexp}, which was interpreted there as a new resonance but shown in \cite{alberto} to be a consequence of the $f_0(17100)$ resonance. 

The approach has proved very solid and highly predictive.  It has also been extended to the charm and beauty sector where some experimental data are well reproduced and the approach leads to predictions of many new resonances \cite{raquelxyz,xiaojuan,liangone,liangtwo}.

\newpage

\subsection{ Review of the $f_0(500)$ properties and its non-ordinary nature from its Regge trajectory}
\addtocontents{toc}{\hspace{2cm}{\sl  J.R.~Pel\'aez}\par}

\vspace{5mm}

 J.R.~Pel\'aez

\vspace{5mm}

\noindent
Departamento de F\'{\i}sica Te\'orica II. Universidad Complutense, 28040 Madrid, SPAIN

\vspace{5mm}

In this talk I first reviewed the recent major revision of the $f_0(500)$ resonance properties in the Particle Data Tables (PDT) \cite{PDG}, and the main results that have driven this change.
After some brief introduction to the history of this controversial state, which is also
known as the $\sigma$ meson, I explained how the combination of new data
with rigorous and model independent approaches has provided very convincing evidence of the existence and properties of this state, which was well known for pratitioners within
the ``scalar meson community'', but has only made it very recently to the PDT, whose approach is  more consensual and conservative. For a recent minireview, see \cite{Pelaez:2013jp}

\begin{figure}[b]
\vspace*{-1.5cm}
  \includegraphics[width=.54\textwidth]{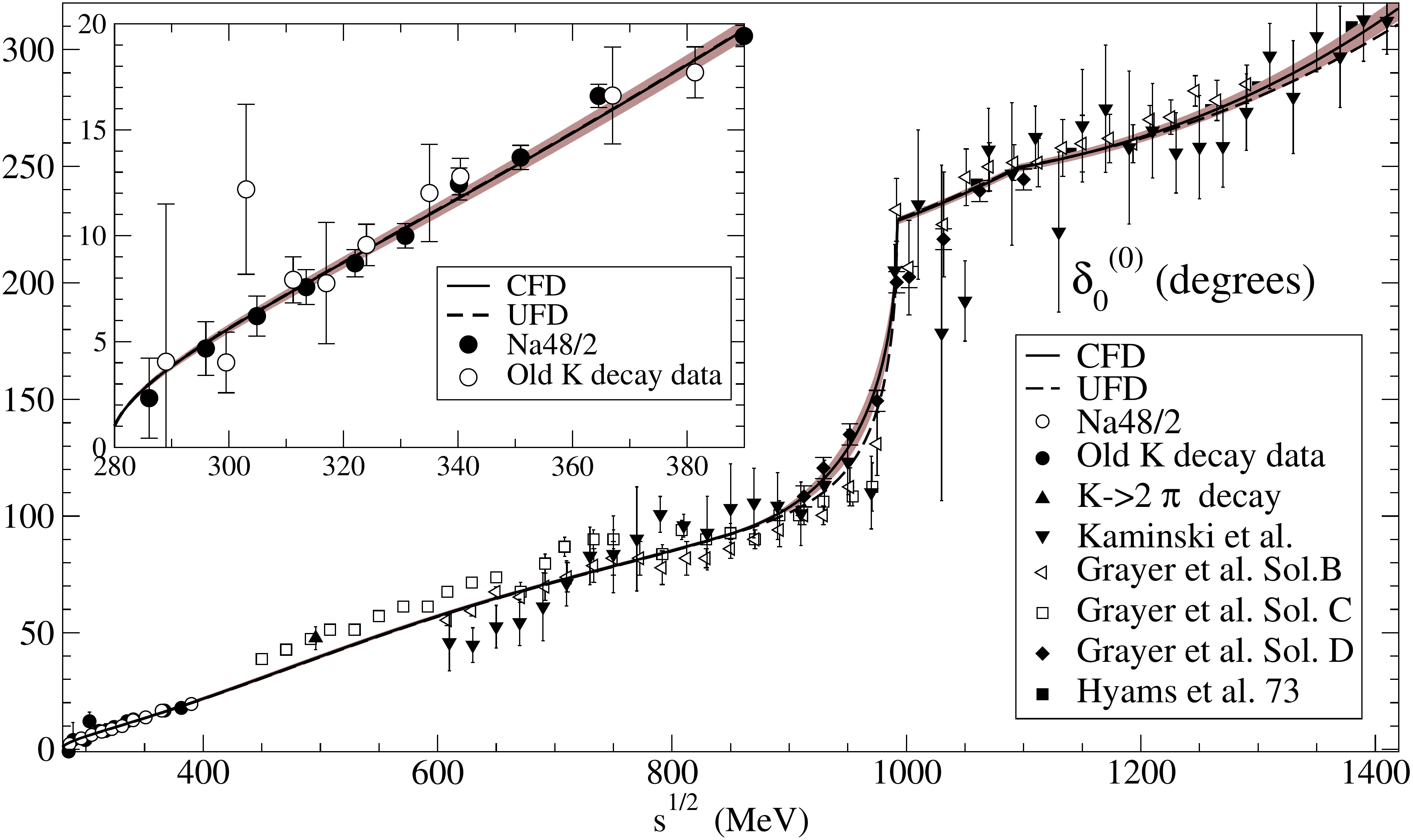}
  \includegraphics[width=.42\textwidth]{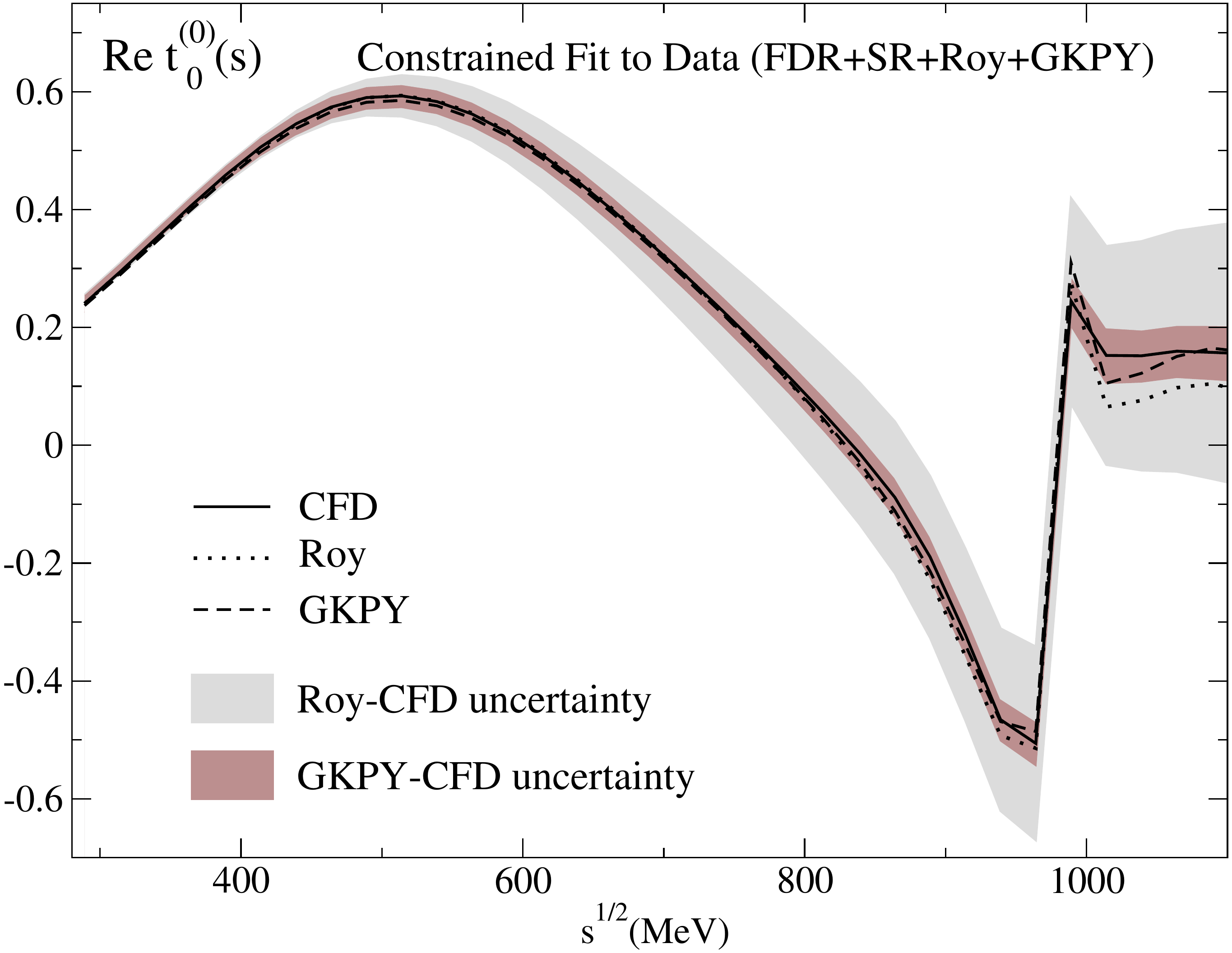}
  \caption{ \small Scalar-isoscalar $\pi\pi$ scattering figures from \cite{GKPY11}. Left: Data on the $\delta_0^{(0)}$ scattering phase versus the dispersive parameterization.
Right: Fulfillment of Roy and GKPY equations for this same wave. 
}
  \label{fig:UFDCFD}
\vspace*{-.5cm}
\end{figure}

An example of the precision attained with dispersive studies is given in Fig.\ref{fig:UFDCFD}, taken from \cite{GKPY11}, where a constrained fit to data was performed. Later on, the dispersion relations are used to obtain the correct analytic continuation to the complex plane in a model independent way, and determine the position and residue of the resonance associated pole.

\begin{figure}[t]
\vspace*{-1.cm}
  \centering
  \includegraphics[width=.6\textwidth]{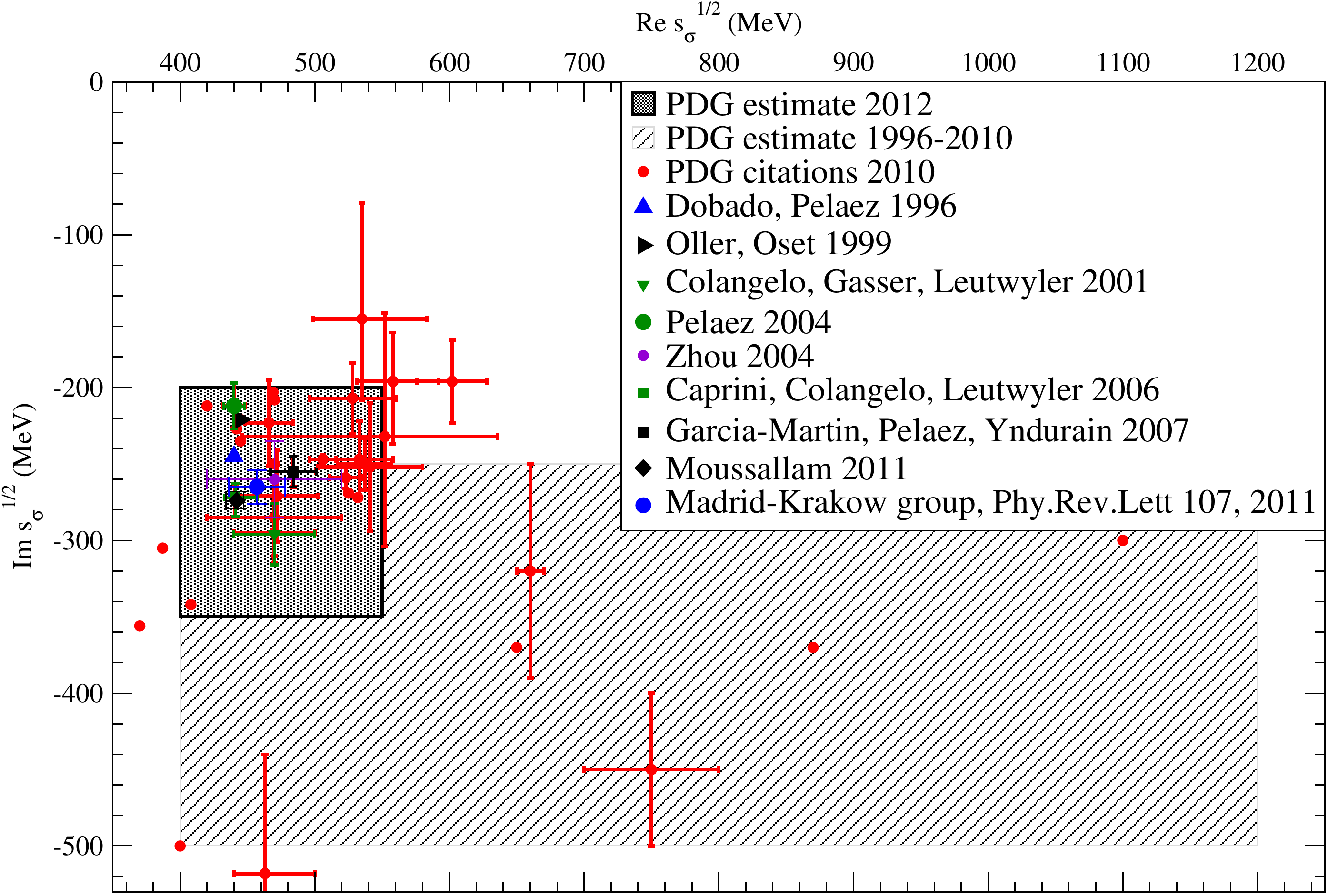}
  \caption{ \small $f_0(500)$ poles in the PDT\protect\cite{PDG}. 
Non-red poles are obtained from dispersive or analytic approaches \protect\cite{others,CGL,Caprini:2005zr,GarciaMartin:2011jx,Moussallam:2011zg}.
}
  \label{fig:poles}
\vspace*{-.5cm}
\end{figure}

As a result of this kind of analyses the 2012 PDT has finally reduced the quoted uncertainties of the $\sigma$ 
mass, by a factor of more than five, down to 400 to 550 MeV,
 and width, by a factor of two, now estimated between 400 and 700 MeV. This new uncertainty estimate is shown in Fig.\ref{fig:poles}, as a dark gray area, versus the old one, represented as a large light gray rectangular area,  which was quoted from the 2002 edition until 2010, despite considering the $\sigma$ meson as a ``well established state''. To my view, these RPP criteria are still rather conservative, and for the $\sigma$ I would only rely on pole extractions based on rigorous analytic methods.  Furthermore, the PDT `Note on light scalars'' suggests that one could 
``take the more radical point of view and just
average the most advanced dispersive analyses'' (here correspond to \cite{CGL,Caprini:2005zr,GarciaMartin:2011jx,Moussallam:2011zg}, shown in Fig.\ref{fig:poles}), to find: $\sqrt{s_\sigma}=(446\pm6)-(276\pm5)\,$ MeV.

\newpage

In this talk I also reported on our recent calculation \cite{Londergan:2013dza} of the Regge trajectory of the $f_0(500)$ meson within a dispersive analysis that allows us to deal with the widths of the resonances. Our only input is the position  and residue of the pole that dominates a given  elastic partial wave in two body scattering. When applied to pions, we obtain an almost real and linear trajectory for the $\rho(770)$ whose intercept and slope is in good agreement with the well known linear trajectories for ordinary hadrons whose slope  is universal and $O(1 {\rm GeV})$. Note that the linear trajectory is not an input, but a result. In contrast, when the same method is applied to the $f_0(500)$ we find a non-real, non-linear trajectory, whose slope at $s=0$ is about two orders of magnitude smaller than the ordinary trajectories. This is a strong hint on the non $q\bar q$ nature of the $f_0(500)$ resonance. 

I thank the organizers for their kind hospitality and the nice workshop organization.

\newpage

\subsection{Pole Identification with Laurent + Pietarinen Expansion in Meson Physics}
\addtocontents{toc}{\hspace{2cm}{\sl  A.~\v{S}varc}\par}

\vspace{5mm}

A.~\v{S}varc

\vspace{5mm}

\noindent
Rudjer Bo\v{s}kovi\'{c} Institute, Bijeni\v{c}ka cesta 54, P.O. Box 180, 10002 Zagreb, Croatia\\

\vspace{5mm}

 We present a new approach to quantifying pole parameters of single-channel processes based on a Laurent expansion of 
partial-wave T-matrices in the vicinity of the real axis \cite{Svarc2013}. Instead of using the conventional power-series 
description of the  non-singular part of the Laurent expansion, we represent this part by a convergent series of Pietarinen 
functions. 
As the analytic structure of the non-singular part is usually very well known (physical cuts with branch points at inelastic 
thresholds, and unphysical cuts in the negative energy plane), we find that one Pietarinen series per cut represents the 
analytic structure fairly reliably. The number of terms in each Pietarinen series is determined by the quality of the fit.  The 
method is tested in two ways: on a toy model constructed from two known poles, various background terms, and two physical cuts, 
and on several sets of realistic $\pi$N elastic energy-dependent partial-wave amplitudes (GWU/SAID - \cite{GWU,GWU1}, and 
Dubna-Mainz-Taipei - \cite{DMT0,DMT}).  We show that the method is robust and confident using up to three Pietarinen series, 
and is particularly  convenient in fits to amplitudes, such as single-energy solutions, coming more directly from experiment; 
cases  where the analytic structure of the regular part is a-priori unknown.

\newpage

\subsection{Light Meson Physics with Crystal Ball at MAMI}
\addtocontents{toc}{\hspace{2cm}{\sl M.~Unverzagt}\par}

\vspace{5mm}

M.~Unverzagt

\vspace{5mm}

\noindent
Institut f\"ur Kernphysik, Johannes Gutenberg-Universt\"at Mainz, Germany\\

\vspace{5mm}

The A2 collaboration at the Institute for Nuclear Physics in Mainz, Germany, carries out experiments with Bremsstrahlung photons derived from electrons in the Glasgow-tagging spectrometer~\cite{McGeorge:2007tg}. The electrons are accelerated in the Mainz Microtron (MAMI)~\cite{Jankowiak:2006yc,Kaiser:2008zza} up to a maximum energy of $E_e = 1604$~MeV. With the Crystal Ball-spectrometer~\cite{Starostin:2001zz} and a forward spectrometer-wall consisting of TAPS-crystals~\cite{Novotny:1991ht} the A2 collaboration performs studies of light meson decays.

Results from the A2 colaboration include the most precise $\eta$ and $\eta'$ photoproduction cross sections to date. The results for the $\eta$ meson cover a wide range from threshold to $\sqrt{s} \approx 1.9$~GeV~\cite{McNicoll:2010qk}. In the case of the $\eta'$ only a limited range can be covered due to the maximum electron energy from MAMI. Nevertheless, a preliminary analysis of the data taken by the A2 collaboration shows unprecedented accuracy in the threshold region.

The A2 collaboration has also studied low-energy QCD in particular $\chi$PT related decays. The isospin-breaking $\eta \to 3\pi^0$, which can be related to the up- and down-quark mass difference, was measured with the worlds best accuracy~\cite{Unverzagt:2008ny,Prakhov:2008ff}. The amplitude of the $\eta \to \pi^0 \gamma \gamma$ decay has first sizable contributions at $O(p^6)$, but the low-energy constants have to be determined from models. Thus, this decay is a stringent test of $\chi$PT at next-to-next-to leading order and also of these models. A soon to be published analysis of this decay gave $1.2 \cdot 10^3$ $\eta \to \pi^0 \gamma \gamma$ events which is currently the most accurate result, but for distinguishing between different models even higher precision has to be reached. The preliminary decay width $\Gamma (\eta \to \pi^0 \gamma \gamma) = (0.33 \pm 0.03_{\mathrm{tot}})$~eV agrees with all theoretical calculations but disagrees with the competitive preliminary result from the KLOE experiment by more than four standard deviations.

The A2 collaboration also contributes to the studies of transition form factors which do not only probe the structure of these particles but might also be of importance for Standard Model calculations of the light-by-light contribution to the Anomalous Magnetic Moment of the Muon. In 2011, the determination of the $\eta$ transition form factor based on $\sim$1350 $\eta \to e^+ e^- \gamma$ events~\cite{Berghauser:2011zz} was published. An independent analysis of 3 times more data gave roughly 20,000 $\eta \to e^+ e^- \gamma$ events. The resulting transition form factor agrees very well with all earlier measurements. Though the result shows good agreement with theoretical calculations the statistical accuracy does not allow for ruling out any prediction. The new result of the A2 collaboration will be published soon.

Breaking of $C$-violation was studied through measuring the branching ratios for the $\omega \to \eta \pi^0$, $\omega \to 2\pi^0$ and $\omega \to 3 \pi^0$ decays~\cite{Starostin:2009zz}. The upper limits determined by the A2 collaboration are to date the only values used by the PDG~\cite{Beringer:1900zz}.

In the next few years the A2 collaboration plans to continue studying the topics mentioned above. The statistics on already analysed decays will be improved greatly. The $\eta/\eta' \to 3\pi^0$ and $\eta' \to \eta \pi^0 \pi^0$ decays will be studied as well as pseudoscalar-vector-$\gamma$ transitions like $\eta' \to \omega \gamma$ and $\omega \to \eta \gamma$. Furthermore, it is planned to investigate transition form factors in $\pi^0/\eta/\eta' \to e^+ e^- \gamma$ and $\omega \to \pi^0 e^+ e^-$ decays. $C$- and $CP$-violation will be examined in $\pi^0/\eta \to 3 \gamma$, $\eta \to 2 \pi^0 \gamma$, $\eta \to 3 \pi^0 \gamma$ and $\eta \to 4\pi^0$ decays. As background for the $\pi^0 \to 3 \gamma$ decay the allowed $\pi^0 \to 4 \gamma$ might be studied which has never been seen yet, but some hadronic models predict a branchiung ratio within the reach of the Crystal Ball at MAMI experiment.

\newpage

\subsection{ Measurements of Kaon Decays}
\addtocontents{toc}{\hspace{2cm}{\sl  R.~Wanke}\par}

\vspace{5mm}

 R.~Wanke

\vspace{5mm}

\noindent
Institut f\"ur Physik, Johannes Gutenberg Universt\"at Mainz, Germany\\

\vspace{5mm}

The NA48 and NA62 experiments at CERN have a long tradition of kaon
decay studies. NA48 as successor of NA31 started measuring direct CP
violation in $K^0$ decays in 1997, followed by NA48/1 in 2002 (rare
$K_S$ and hyperon decays) and NA48/2 in 2003 and 2004 ($K^\pm$
decays). 
In the year 2007, still with the original NA48 detector, a long
data-taking period was performed by the already formed NA62
collaboration for the precise measurement of the ratio 
$R_K = \Gamma(K \to e \nu)/\Gamma(K \to \mu \nu)$.

Here we report on recent measurements of NA48/2 and NA62 ($R_K$ phase)
on rare kaon decays and on prospects for the future NA62 experiment,
which starts data-taking with a new detector at the end of 2014.

\paragraph{Precise Measurement of $K^\pm \to \pi^\pm \gamma \gamma$}

\begin{figure}[h]
  \includegraphics[width=0.302\linewidth]{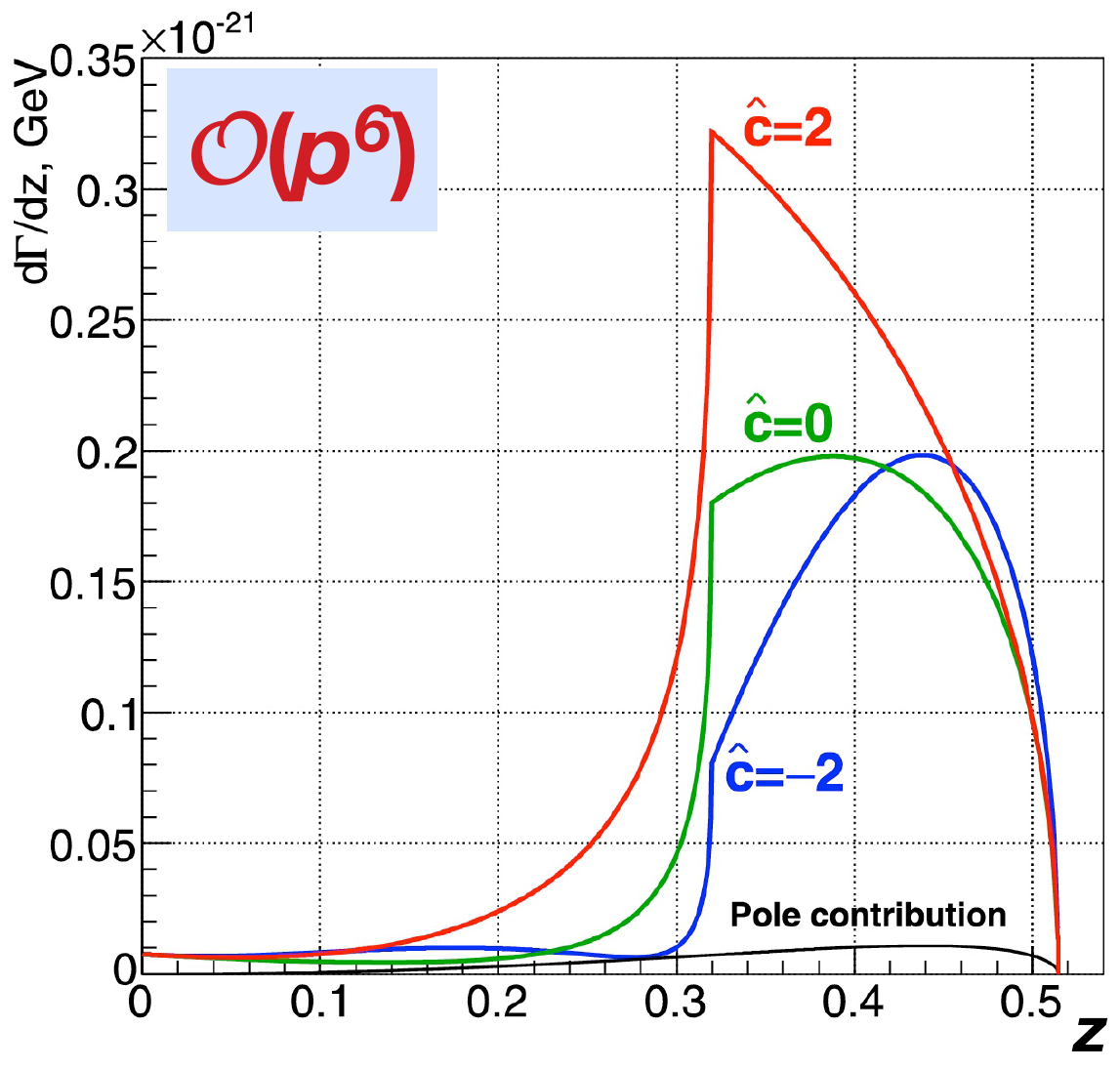}
  \hspace*{\fill}
  \includegraphics[width=0.29\linewidth]{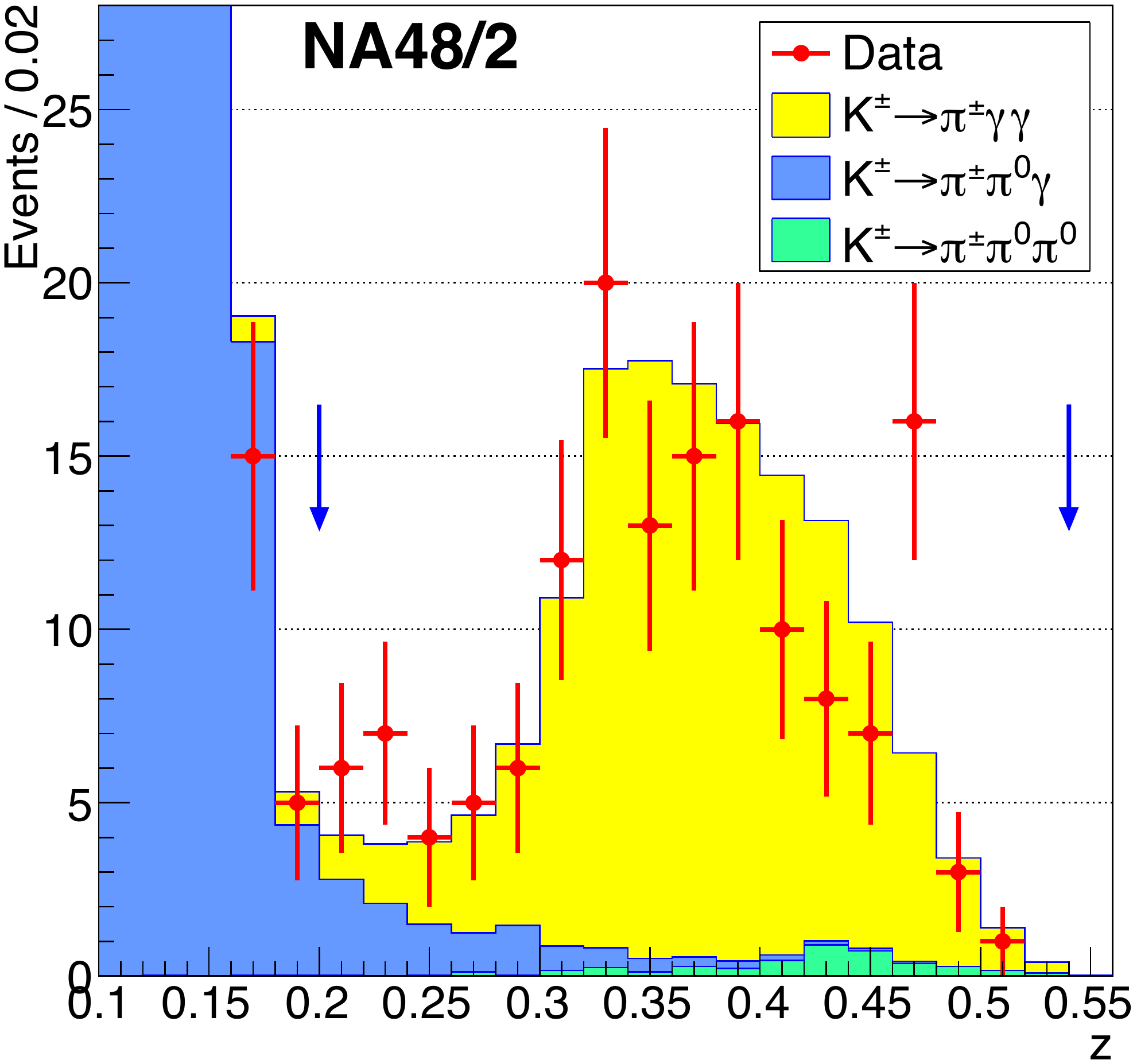}
  \hspace*{\fill}
  \includegraphics[width=0.29\linewidth]{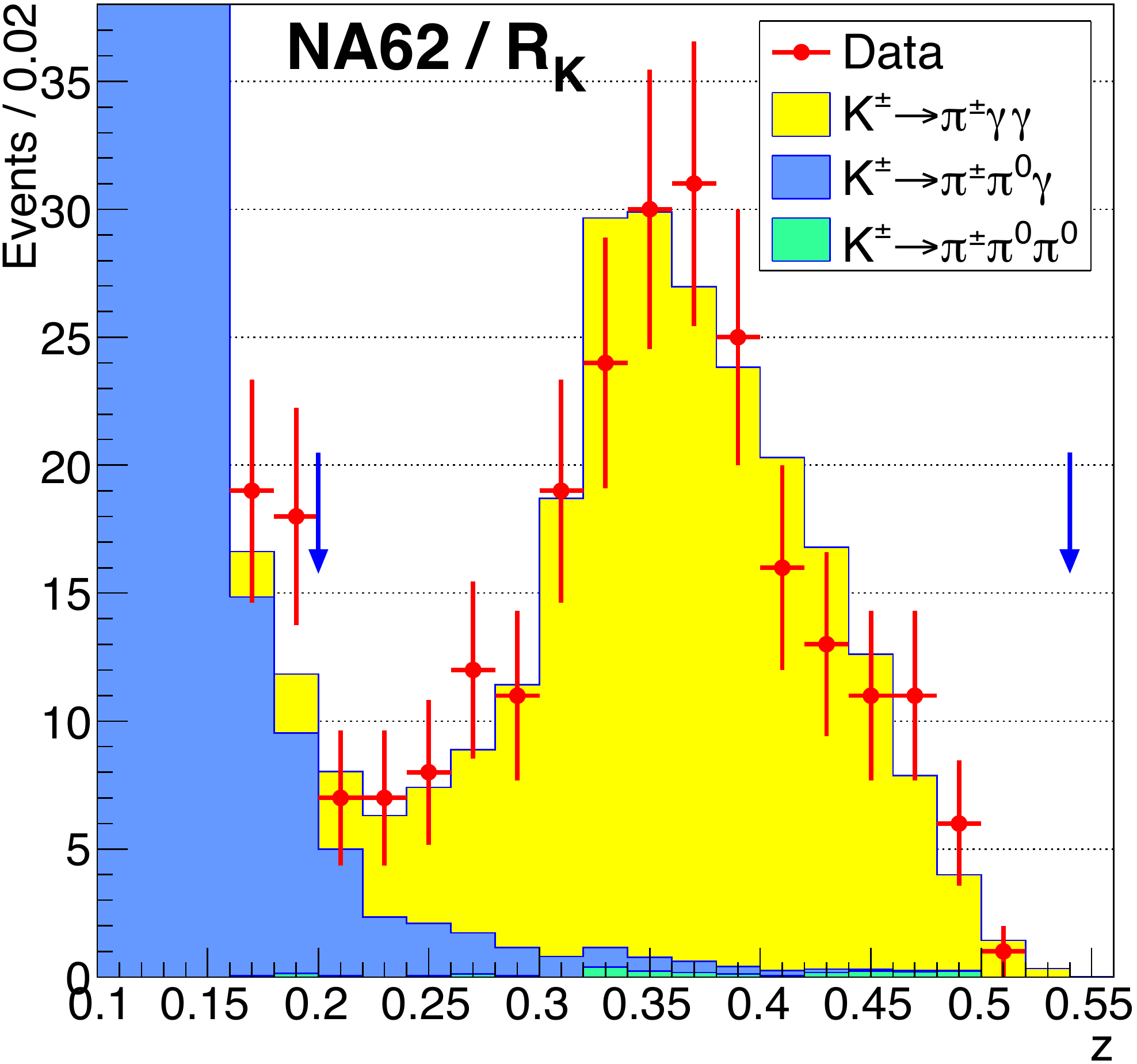}
  \caption{Distributions of $z=m_{\gamma\gamma}^2/m_K^2$ for
    ${\cal O}(p^6)$ ChPT (left) and NA48/2~\cite{NA62:2014oza} (center) and 
    NA62/$R_K$-phase data~\cite{NA62:2014oza} (right). The arrows indicate the
    region used for the $\hat{c}$ extraction.}
  \label{fig:Kpigg_z}
\end{figure}

The amplitudes of $K \to \pi \gamma \gamma$ decays have no
contributions of ${\cal O}(p^2)$ in Chiral Perturbation Theory
(ChPT). Moreover, at ${\cal O}(p^4)$ only two-pion loop
diagrams contribute, resulting in a Wigner-cusp at $2\,m_{\pi^+}$ in the
$\gamma \gamma$ invariant mass. For the charged
decay $K^\pm \to \pi^\pm \gamma \gamma$, the ${\cal O}(p^4)$ amplitude
depends on only one free parameter $\hat{c}$ which should be of 
${\cal O}(1)$~\cite{Ecker:1987hd}. At the following ${\cal O}(p^6)$
additional contributions as unitarity corrections and pole
contributions have to be taken into
account~\cite{DAmbrosio:1996id}, resulting e.g. in a
non-zero rate at $z=0$ as shown in Fig.~\ref{fig:Kpigg_z}~(left).

Both NA48/2 and NA62 have collected comparable data samples of
$K^\pm \to \pi^\pm \gamma \gamma$ decays, resulting in a total of 324
candidates with an expected background of $27.9 \pm 1.3$ events from
$K^\pm \to \pi^\pm \pi^0 \gamma$ and $K^\pm \to \pi^\pm \pi^0 \pi^0$
decays. For both data sets separate analyses were
undertaken~\cite{NA62:2014oza,Batley:2014fy}.
Combining both measurements yields (in ${\cal O}(p^6)$ ChPT) a value
of $\hat{c} = 2.00 \pm 0.26$, where the uncertainty is dominated by the
data statistics.

\paragraph{First Observation of $K^\pm \to \pi^\pm \pi^0 e^+ e^-$}

The decay $K^\pm \to \pi^\pm \pi^0 e^+ e^-$ is similar to 
$K^\pm \to \pi^\pm \pi^0 \gamma$ with an internal photon conversion.
It is dominated by inner bremsstrahlung (IB) while direct photon
emission (DE) is a sub-leading effect of ${\cal O}(p^4)$
ChPT~\cite{Pichl:2001is}. 

\begin{figure}
  \begin{minipage}[t]{0.3\linewidth}
    \centering
    \includegraphics[width=\linewidth]{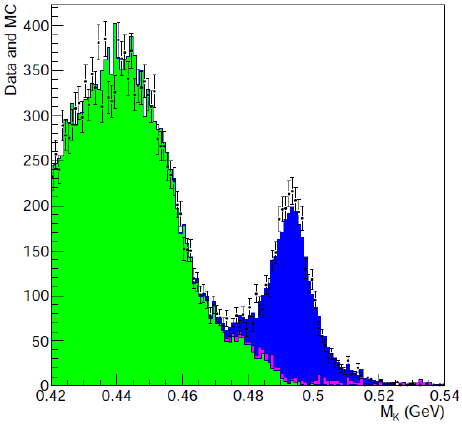}
    \caption{$K^\pm \to \pi^\pm \pi^0 e^+ e^-$ signal (blue) for 
      $\approx 40\,\%$ of the NA48/2 data (in green: 
      $K^\pm \to \pi^\pm \pi^0 \gamma$)~\cite{FB:HEPMAD13}.}    
    \label{fig:Kpipiee_mass}
  \end{minipage}
  \hspace*{\fill}
  \begin{minipage}[t]{0.3\linewidth}
    \centering
    \includegraphics[width=0.895\linewidth]{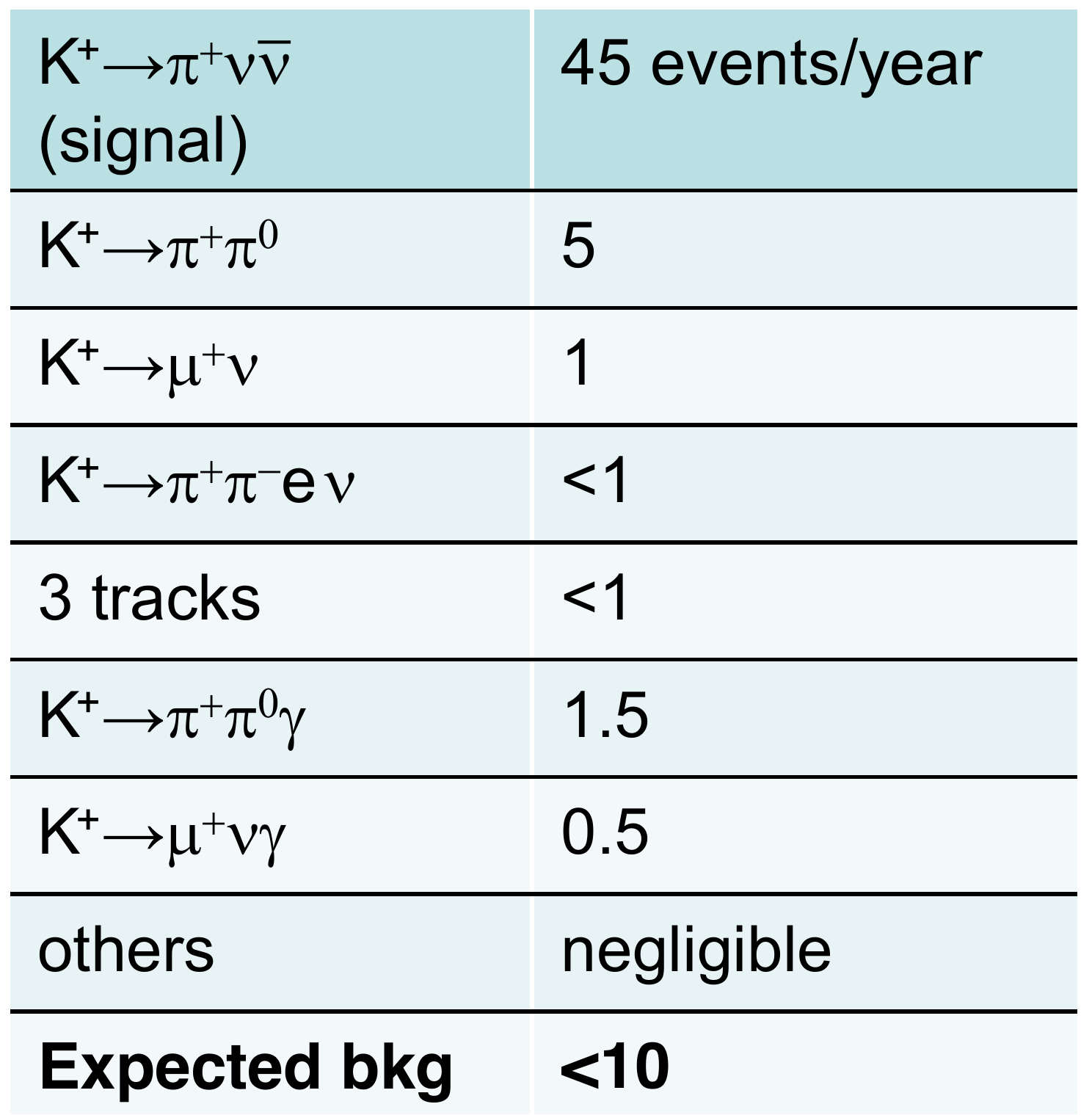}
    \caption{NA62 expectations for 
      $K^+ \to \pi^+ \nu \bar{\nu}$ SM-events and background for one year
      of data-taking~\cite{EG:MesonNet2013}.} 
    \label{fig:Kpinunu_yield}
  \end{minipage}
  \hspace*{\fill}
  \begin{minipage}[t]{0.3\linewidth}
    \includegraphics[width=0.963\linewidth]{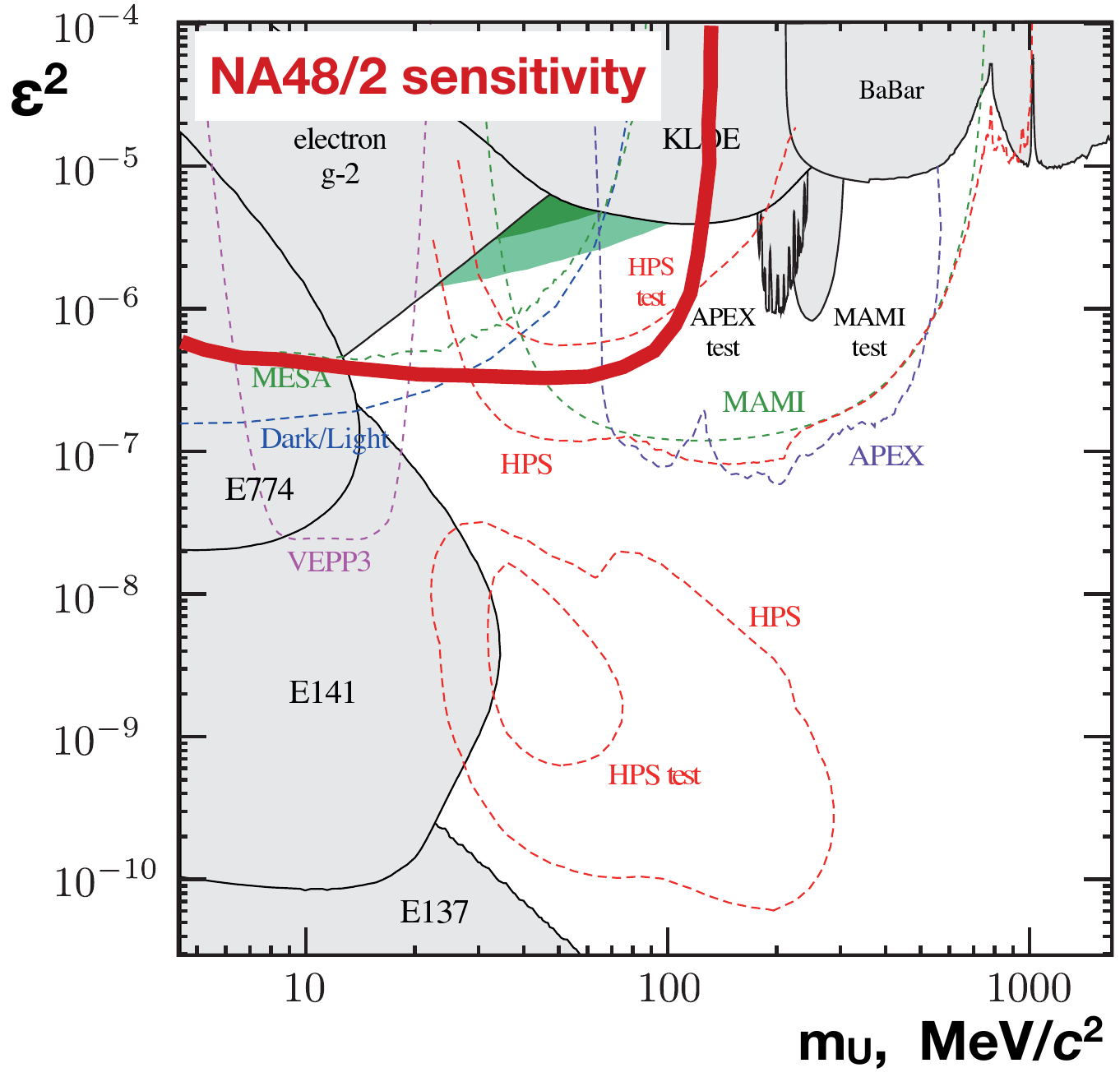}
    \caption{Various exclusion limits for the $U$
      boson~\cite{Endo:2012ca}. The sensitivity of NA48/2 is
      indicated in red~\cite{EG:MesonNet2013}.}
    \label{fig:UReach_NA48}
  \end{minipage}
\end{figure}

Using about $40\,\%$ of their recorded data, NA48/2 has now
reported the first observation of the decay 
$K^\pm \to \pi^\pm \pi^0 e^+ e^-$ with about 2500 signal candidates
and an estimated background of 280 events
(Fig.~\ref{fig:Kpipiee_mass}) \cite{FB:HEPMAD13}. The analysis of the
data is on-going.

\paragraph{Future Reach for rare Kaon Decays}

The aim of the new NA62 experiment is the measurement of about
100 Standard Model (SM) events of the decay $K^+ \to \pi^+ \nu \bar{\nu}$
in two years of data taking (Fig.~\ref{fig:Kpinunu_yield}). With this
statistical precision a huge amount of possible New Physics scenarios
can either be found or ruled out.

In addition, the expected unprecedented statistics on $K^+$ decays
will allow to search for a variety of rare, forbidden, and non-SM
$K^+$ decays. An example is the search of the so-called dark photon
or $U$ boson, where already the on-going analysis of NA48/2 data will
significantly improve the existing limits (see
Fig.~\ref{fig:UReach_NA48})~\cite{EG:MesonNet2013}.

\newpage

\subsection{ Meson Production and Decays with WASA-at-COSY}
\addtocontents{toc}{\hspace{2cm}{\sl M.~Wolke}\par}

\vspace{5mm}

 M.~Wolke \footnote{Supported by the Swedish Research 
 Council (VR).} for the WASA-at-COSY Collaboration

\vspace{5mm}

\noindent
Department of Physics and Astronomy, Uppsala University, Sweden\\

\vspace{5mm}

New results have been obtained with the WASA detector at COSY on the light 
meson decays $\pi^0 \rightarrow e^+e^-\gamma$ and $\eta \rightarrow 
\pi^+\pi^-\pi^0$ as well as on elastic neutron proton scattering in the energy 
range of a narrow resonance--like structure observed in double pionic fusion 
reactions.

Speculatively, recent unanticipated astrophysical observations (for refs.\ see 
e.\ g.~\cite{Archilli:2011zc,Adlarson:2013eza}) might be explained by a 
dark matter WIMP from a secluded gauge sector under which SM particles are 
uncharged (see~\cite{Essig:2013lka} for a review, and~\cite{Fayet:1980ad} for 
early incarnations of the concept of a new light boson).
In one scenario, the coupling to the SM arises from the kinetic mixing of the 
gauge boson (with a mass at the GeV scale) of a dark 
$\mbox{U}\left(1\right)_d$ with the SM $\mbox{U}\left(1\right)$, with the 
mixing parameter $\epsilon$ expected of the order of $10^{-4} - 10^{-2}$.
Due to its small width the $U$ boson or {\it dark photon} should be observable 
as a narrow peak in the lepton--antilepton invariant mass in light meson 
conversion decays.
Search channels include $\phi \rightarrow \eta U$, $\eta \rightarrow 
\gamma U$, and $\pi^0 \rightarrow \gamma U$ with $U \rightarrow e^+e^-$.
New results from WASA-at-COSY on the decay $\pi^0 \rightarrow 
\gamma e^+e^-$~\cite{Adlarson:2013eza} constrain the parameter space for $U$ 
boson masses $20\,\mbox{MeV} \le \mbox{m}(U) \le \mbox{100}\,\mbox{MeV}$ and 
mixing parameter values around $2 - 3 \cdot 10^{-3}$ further compared to 
previous results in this range by SINDRUM~\cite{MeijerDrees:1992kd} and 
KLOE~\cite{Archilli:2011zc}.
An intriguing motivation to study particularly this part of the parameter 
space is the possible $\mbox{U}\left(1\right)_d$ contribution to the anomalous 
magnetic moment of the muon. 
To get agreement between theory and experiment for the muon anomaly within two 
standard deviations defines a {\it welcome band}~\cite{Pospelov:2008zw}.
For the mass range given above, we expect the full WASA-at-COSY statistics, 
which is at least an order of magnitude larger, to cover this band completely.

The decays $\eta \rightarrow 3\,\pi$ proceed via isospin symmetry breaking in 
the strong interaction, and electromagnetic corrections are expected to be 
small~\cite{Ditsche:2008cq}.
A precision determination of the light quark mass difference requires an 
accurate theoretical calculation, reproducing the dynamics of the $3\,\pi$ 
final state (see e.\ g.~\cite{Kampf:2011wr} and references therein).
Theoretical predictions can be tested with precision data on the $\eta 
\rightarrow 3\,\pi$ Dalitz plots.
The largest statisitics presently available are from the KLOE 
experiment with the final Dalitz plot containing $1.3 \cdot 10^6$ 
events~\cite{Ambrosino:2008ht}.
However, the parameters $a$ and $b$ in the obtained Dalitz plot 
parametrisation~\cite{Ambrosino:2008ht} are difficult to reproduce 
theoretically.
An independent measurement has been done with WASA-at-COSY using the tagging 
reaction $p d \rightarrow \mbox{}^3\mbox{He}\,\eta$ and a sample of 
$1.7 \cdot 10^5$ $\eta \rightarrow \pi^+ \pi^- \pi^0$ events.
Within $2\,\sigma$ the WASA-at-COSY values~\cite{Adlarson:2014zz} confirm 
both the published KLOE data~\cite{Ambrosino:2008ht} as well as preliminary 
results on the full KLOE statistics.
A significantly larger sample of $\eta$ decays has been measured with the 
WASA detector in $pp \rightarrow pp \eta$, and statistics are expected to 
be comparable to the data from~\cite{Ambrosino:2008ht}.

Recently, a narrow resonance--like structure has been observed in the 
elementary double pionic fusion reactions $pn \rightarrow d \pi^0 \pi^0$ and 
$pn \rightarrow d \pi^+ \pi^-$~\cite{Bashkanov:2008ih}.
With no conventional explanation at hand as of now, the signal is consistent 
with an s--channel resonance in the proton--neutron and $\Delta$--$\Delta$ 
systems with a mass of $2380\,\mbox{MeV}/\mbox{c}^2$, a width of 
$\Gamma \approx 70\,\mbox{MeV}$ and quantum numbers 
$I \left( J^P\right) = 0 \left( 3^+\right)$ favoured by the deuteron and pion 
angular distributions.
Such a resonance must also be observable in elastic proton--neutron 
scattering, in particular in the analysing power $A_y$ which is determined 
only by interference terms of the partial waves contributing.
Since there have been no experimental data on the $n\,p$ analysing power in 
the energy range of the resonance so far, data have been taken with the WASA 
detector in the quasifree mode, $\vec{d} p \rightarrow p n + p_{spectator}$.
Measured analysing powers have been included in the SAID database, and a new 
partial wave analysis has been performed, which shows a resonance pole in the 
coupled $\mbox{}^3 D_3 - \mbox{}^3 G_3$ partial waves as expected from the 
resonant structure observed in the double pionic fusion 
reactions~\cite{Adlarson:2014pxj}.
Such a resonance might be interpreted as a hidden--colour six--quark 
state, but is also reproduced in recent quark model calculations as well 
as using a purely hadronic model for pions, nucleons, and 
$\Delta$'s~\cite{Bashkanov:2013cla}.

\newpage

\section{List of participants}

\begin{flushleft}
\begin{itemize}
\item Patricia Bickert, Universt\"at Mainz, {\tt bickert@kph.uni-mainz.de}
\item Johan Bijnens, Lund University, {\tt  bijnens@thep.lu.se}
\item Nikolay Borisov, JINR
\item Sasa Ceci, Rudjer Boskovic Institue, Zabreg, {\tt ceci@irb.hr}
\item Rafel Escribano, Universitat Automona de Barcelona, {\tt rescriba@ifae.es }
\item Shuangshi Fang, IHEP Beijing, {\tt fangss@ihep.ac.cn}
\item Simona Giovannella, Laboratori Nazionali di Frascati, {\tt  simona.giovannella@lnf.infn.it}
\item Wolfgang Gradl, Universit\"at Mainz, {\tt gradl@kph.uni-mainz.de }
\item Mirza Hadzimehmedovic, University of Tuzla, {\tt mirza.hadzimehmedovic@untz.ba}
\item Christoph Hanhart, Forschungszentrum J\"ulich, {\tt c.hanhart@fz-juelich.de}
\item Karol Kampf, Charles University, {\tt karol.kampf@mff.cuni.cz}
\item Mari\'an Koles\'ar, Charles University, {\tt  kolesar@ipnp.troja.mff.cuni.cz}
\item Bastian Kubis, University of Bonn, {\tt kubis@hiskp.uni-bonn.de}
\item Andrzej Kupsc, Uppsala University, {\tt Andrzej.Kupsc@physics.uu.se }
\item Alezander Lazarev, JINR, Dubna
\item Stefan Leupold, Uppsala University, {\tt stefan.leupold@physics.uu.se}
\item Matthias F.M. Lutz, GSI Darmstadt, {\tt  m.lutz@gsi.de}
\item Lefteris Markou, the Cyprys Institute, {\tt l.markou@cyi.ac.cy }
\item Pere Masjuan, Universt\"at Mainz, {\tt masjuan@kph.uni-mainz.de}
\item Bachir  Moussallam, IPN, Universit\'e Paris-Sud XI, {\tt moussall@ipno.in2p3.fr}
\item Andreas Neiser, Universit\"at Mainz, {\tt neiser@kph.uni-mainz.de }
\item Eulogio Oset, University of Valencia, {\tt oset@ific.uv.es }
\item Hedim Osmanovic, University of Tuzla, {\tt hedim.osmanovic@untz.ba }
\item Michael Ostrick, Universit\"at Mainz, {\tt ostrick@kph.uni-mainz.de }
\item Jose Pelaez, Universidad Complutense de Madrid, {\tt jrpelaez@fis.ucm.es }
\item Sergey Prakhov, UCLA/Universit\"at Mainz, {\tt prakhpv@ucla.edu}
\item Christoph Florian Redmer, Universit\"at Mainz, {\tt redmer@uni-mainz.de }
\item Stefan Scherer, Universit\"at Mainz, {\tt scherer@kph.uni-mainz.de}
\item Jugoslav Stahov, University of Tuzla, {\tt jugoslav.stahov@untz.ba }
\item Oliver Steffen, Universit\"at Mainz, {\tt steffen@kph.uni-mainz.de}
\item Alfred \v{S}varc, Rudjer Boskovic Institue, Zagreb, {\tt svarc@irb.hr }
\item Carla Terschluesen, Uppsala University, {\tt carla.terschluesen@physics.uu.se}
\item Andreas Thomas, Universit\"at Mainz, {\tt thomas@kph.uni-mainz.de }
\item Lothar Tiator, Universit\"at Mainz, {\tt tiator@kph.uni-mainz.de}
\item Marc Unverzagt, Universit\"at Mainz, {\tt unvemarc@kph.uni-mainz.de }
\item Sascha Wagner, Universit\"at Mainz, {\tt sascha.wagner.89@gmail.com }
\item Rainer Wanke, Universit\"at Mainz, {\tt Rainer.Wanke@uni-mainz.de }
\item Martin Wolfes, Universit\"at Mainz, {\tt wolfes@kph.uni-mainz.de }
\item Magnus Wolke, Uppsala University, {\tt magnus.wolke@fysast.uu.se}
\end{itemize}
\end{flushleft}

\end{document}